\documentclass[prb,twocolumn,superscriptaddress,floatfix,longbibliography]{revtex4-2}
\usepackage{amssymb}
\usepackage{physics}
\usepackage{graphicx}
\usepackage{siunitx}
\usepackage{times}
\usepackage{amsmath}
\usepackage{amsthm}
\usepackage{amsfonts}
\usepackage[T1]{fontenc}
\usepackage{array}
\usepackage{multirow}
\usepackage{color}
\usepackage{esint}
\usepackage{bm}
\usepackage{xcolor}
\usepackage{bbm}
\usepackage[english]{babel}

\usepackage{float}
\usepackage[section]{placeins}

\usepackage{hyperref}
\hypersetup
{	colorlinks,%
	citecolor=green,%
	linkcolor=red,%
	urlcolor=blue%
}
\allowdisplaybreaks[4]

\begin{document}

\title{Interplay of disorder and interaction in quantum Hall systems: from fractional quantum Hall liquids to Wigner crystals and amorphous solids}

\author{Ke Huang}
\affiliation{Department of Physics, City University of Hong Kong, Kowloon, Hong Kong}

\author{Sankar Das Sarma}
\affiliation{Condensed Matter Theory Center and Joint Quantum Institute,
University of Maryland, College Park, Maryland 20742, USA}

\author{Xiao Li}
\affiliation{Department of Physics, City University of Hong Kong, Kowloon, Hong Kong}
\email{xiao.li@cityu.edu.hk}

\date{\today}

\begin{abstract}
We investigate the interplay of disorder and interaction in two-dimensional electron systems in a strong magnetic field, focusing on the transition between Wigner crystals and fractional quantum Hall liquids. 
We first study classical Wigner crystals with charged impurities, revealing an evolution from a coherent crystal to local crystalline domains with short-range order and eventually to an amorphous state as impurity concentration increases. 
We then analyze noninteracting quantum electron crystals created by periodic potentials, showing that their structure factor exhibits both peaks and rings, distinct from classical Wigner crystals. 
Finally, we explore fractional quantum Hall liquids with random short-range disorder and quenched charged impurities, demonstrating that the ground state can evolve from an incompressible liquid to a localized ordered state and eventually to an amorphous state as disorder strength increases. 
In general, we find that random charged impurities lead to longer-range crystalline ordering than the short-range random disorder. 
Our findings highlight the rich interplay between disorder and interaction in quantum Hall systems and provide insights into experimental observations of these phenomena. 
By qualitative comparison with a recent STM experiment [Nature \textbf{628}, 287 (2024)], we conclude that the 2D system crosses over from an incompressible homogeneous fractional quantum Hall liquid to a generic locally ordered solid and eventually to a disordered amorphous solid at large disorder. 
\end{abstract}

\maketitle

\section{Introduction}
Two-dimensional electron systems (2DES) in a strong magnetic field have been a fertile ground for exploring the interplay between disorder and interaction, leading to a rich variety of quantum phases~\cite{Tsui1982,Stormer1999,Prange1990,DasSarma1997,Halperin2020}. 
Since the magnetic field quenches the lowest Landau level kinetic energy in the high-field limit, the emergent quantum phases are all by definition strongly correlated. 
Among these correlated phases, Wigner crystals (WCs)~\cite{Wigner1934} and fractional quantum Hall (FQH) liquids (FQHLs)~\cite{Laughlin1983} represent two outstanding examples of interaction-driven strongly correlated states that emerge from the same underlying physics but exhibit drastically different properties.
In particular, both phases arise from electron-electron interaction in partially filled Landau levels: at low filling factors, the Coulomb energy dominates and electrons crystallize into a WC, while at specific rational fillings, strong correlations stabilize incompressible FQHLs with topological order.
The generic background phase is the Wigner crystal since the kinetic energy is completely quenched, but Laughlin established~\cite{Laughlin1983} that at special filling factors, e.g., the $1/3$ filling where the FQH effect is seen, the system optimizes its energy by forming a homogeneous incompressible liquid, leading to cusps in the energy as a function of filling stabilizing the FQH phase over the WC phase~\cite{Yoshioka1983,Lam1984,Levesque1984,Zhu1993,Price1995,Yi1998,Yang2001}. 
It is interesting to note that the seminal experimental discovery of fractional quantum Hall effects (FQHE) was actually a search for the WC with the abstract stating: ``The formation of a Wigner solid or charge-density-wave state with triangular symmetry is suggested as a possible explanation.''~\cite{Tsui1982} 
It is ironic that the original guess for the FQHE was a Wigner crystal ground state although the actual ground state turns out to be an incompressible quantum liquid stabilized by interactions only at a discrete set of fractional filling factors. 
The currently accepted theoretical picture is that the pristine 2D strong-field system is indeed a WC at generic Landau level fillings except that the incompressible FQHL is energetically stabilized at a series of fractional fillings (which is a measure zero set in the pristine system).

In a strong magnetic field, the quenching of kinetic energy indeed favors Wigner crystallization, and early transport experiments in GaAs heterostructures provided evidence for a magnetic-field-induced WC competing with FQH states~\cite{Andrei1988,Goldman1990,Jiang1990}. 
However, disorder is inevitable in real materials and can profoundly affect both WCs and FQHLs; consequently, transport cannot distinguish between a disorder-pinned WC and a disorder-induced strongly localized amorphous solid phase, since both manifest insulating localized behavior~\cite{Ahn2023,Babbar2026,Huang2026}. 
Most recently, scanning tunneling microscopy (STM) of high-quality bilayer graphene has imaged the WC lattice in real space~\cite{Tsui2024}, furnishing the most compelling direct evidence to date for Wigner crystallization in quantum Hall systems. 

\begin{figure*}[t]
	\center
	\includegraphics[width=\textwidth]{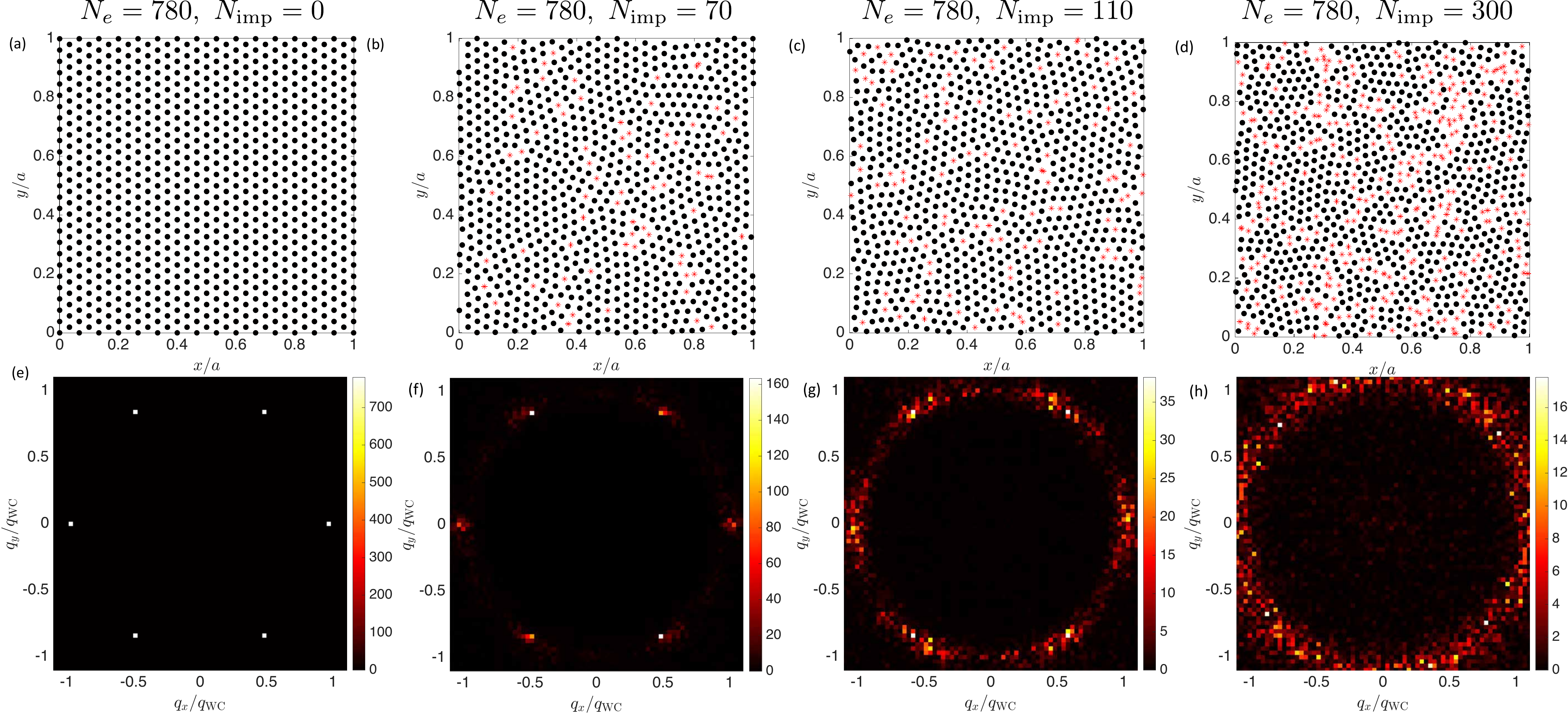}
	\caption{\label{Fig:ClassicalWC} 
	Real-space distribution (top panels) and structure factor (bottom panels) for classical WC with $N_e=780$ electrons and $N_{\text{imp}}=0,70,110,300$ impurities, respectively. In (a-d), the black dots are electrons, and the red stars are impurities with charge $Z=1$. The real-space coordinates in (a)-(d) are scaled by the lattice constant $a$, and the momenta in (e-h) are scaled by the length of the reciprocal vector of the pristine WC, given by $q_{\text{WC}}=\frac{2\pi}{a}\sqrt{\frac{2N_e}{\sqrt3}}$.
	}
\end{figure*}

Previous studies of disorder in quantum Hall systems have focused mainly on the transition from an incompressible FQHL to a localized state as disorder strength increases~\cite{Sheng2003}. 
However, a systematic study that places WCs and FQHLs on equal footing, including nonperturbative disorder effects in the context of the compelling STM experiment~\cite{Tsui2024}, tracing how disorder drives each phase through a common sequence of structural transformations, has never been carried out. 
In particular, the role of being slightly away from the precise FQH filling in this context remains unexplored. 
This question has acquired new urgency with the recent STM imaging of WCs in bilayer graphene~\cite{Tsui2024}, which revealed not only pristine crystalline order but also arc-like amorphous structures whose microscopic origin remains unclear. 
Key open questions include: how disorder fragments a coherent crystal into local domains and eventually into an amorphous state, how different types of disorder (i.e., random potentials versus charged impurities) govern the resulting correlation lengths, how the precise fractional filling affects the physics, and whether finite-temperature thermal fluctuations can restore an FQHL from a disorder-pinned WC. 
These are important questions, central to not only understanding the recent STM experiment~\cite{Tsui2024}, but also the many earlier experiments claiming the manifestation of WC in the lowest Landau level. 
Our work encompasses FQHE, WC, and amorphous solids in the presence of strong disorder and strong interaction within one unified nonperturbative framework. 

In this work, we address these questions by combining classical energy minimization with exact diagonalization of quantum Hamiltonians in the lowest Landau level. 
We first study classical Wigner crystals with charged impurities, establishing a benchmark for how disorder fragments a crystal into local domains and ultimately into an amorphous state. 
We then analyze noninteracting quantum electron crystals created by periodic potentials, revealing that their projected structure factor carries both Bragg-like peaks and a ring of quantum-mechanical origin, distinct from the classical case. 
Turning to the interacting problem, we investigate FQHLs subject to both random short-range disorder and long-range Coulomb impurities, demonstrating that the ground state undergoes a progression from an incompressible liquid to a localized state with short-range crystalline order, and eventually to an amorphous state, with the latter bearing a striking resemblance to the arc-like structures observed experimentally~\cite{Tsui2024}. 
Finally, we show that at finite temperatures, thermal ionization of impurity-bound electrons can restore itinerant carriers and drive a crossover from a pinned Wigner crystal back to an FQH state with free quasiholes.

We present our main results in the following sections of the main text, with more detailed results in the Appendix. 

\section{Classical Wigner crystals with charged impurities}

\subsection{Classical Wigner crystals}
We begin our analysis by studying classical WCs in the presence of random charge impurities as a benchmark for the quantum case. We focus on the classical system at zero temperature, where the electrons' kinetic energy is completely quenched. Hence, the energy of the system is composed of the electron-electron and electron-impurity Coulomb interaction, given by
\begin{align}
	E=\frac12\sum_{i\neq j}\frac{1}{\abs{\vb r_i-\vb r_j}}+Z\sum_{i=1}^{N_e}\sum_{j=1}^{N_{\text{imp}}}\frac{1}{\abs*{\vb r_i-\vb r_j^{\text{imp}}}},
\end{align}
where $\vb r_i$ is the position of the $i$th electron, and $\vb r_i^{\text{imp}}$ is the position of the $i$th impurity with charge $Z$. 
In the absence of impurities, the electrons form a hexagonal lattice in the thermodynamic limit but suffer from finite-size effects. 
For open boundary conditions, electrons accumulate at the boundaries and significantly deviate from the hexagonal lattice. For periodic boundary conditions (PBCs), the finite-size effects appear as defects, which can be dramatically reduced when the WC is nearly commensurate with the PBC. In this work, we study a square torus of length $a$, where a commensurate structure is allowed when $N_e=2xy$ for two integers $x,y$ satisfying $x/y\approx\sqrt{3}$. 
In Fig.~\ref{Fig:ClassicalWC}(a), we show the commensurate structure for $x=26$ and $y=15$. 

\begin{figure*}[!tbp]
	\center
	\includegraphics[width=0.85\textwidth]{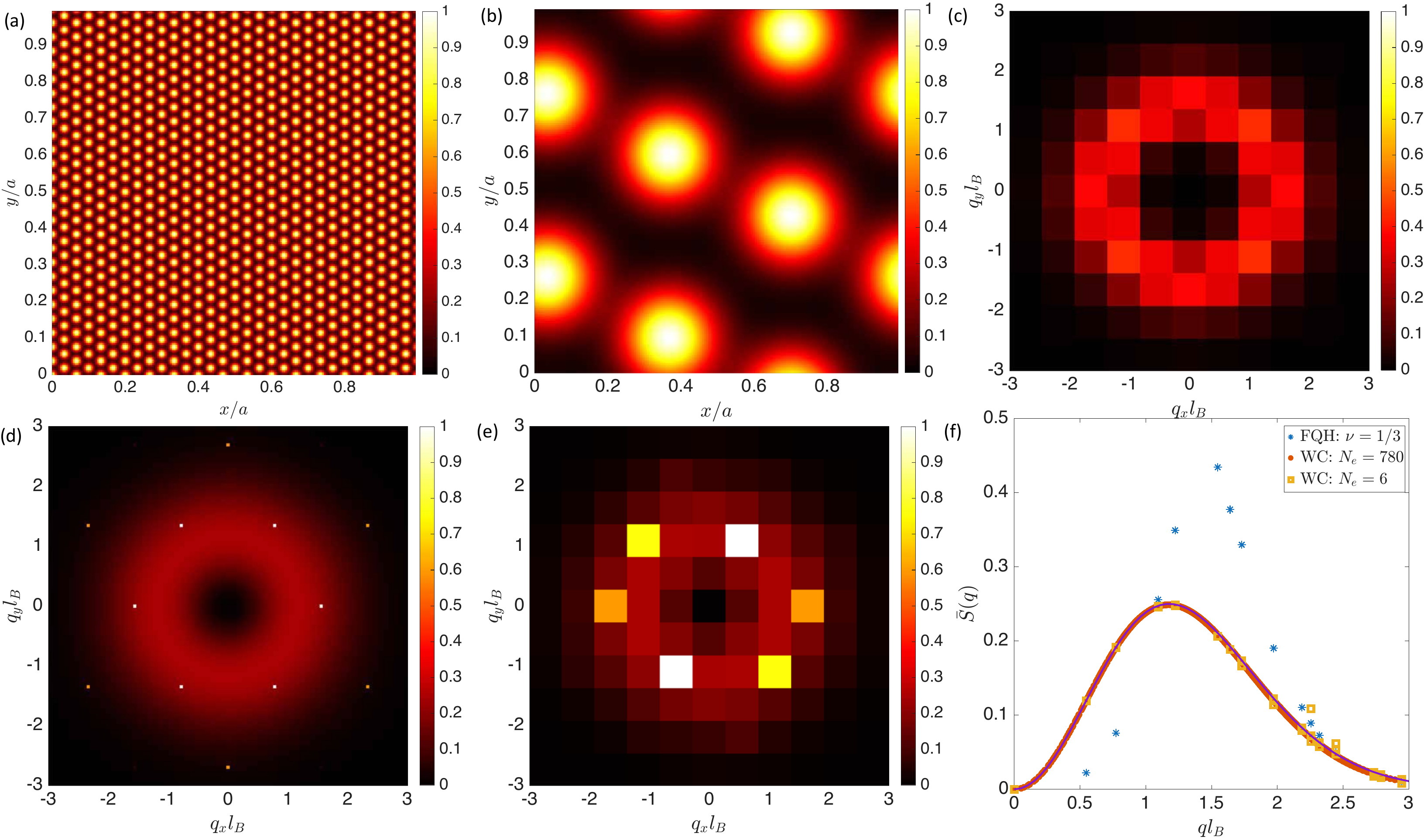}
	\caption{\label{Fig:Compare} 
	(a) Normalized density $2\pi l_B^2 n(r)$ of a WC of $N_e=780$ at $\nu=1/3$. 
	(b) Normalized density of a WC of $N_e=6$ and $N_\phi=21$. 
	(c)-(e) are the projected structure factors of the FQH state of $N_e=7$ at $\nu=1/3$, the electron crystal of $N_e=780$ at $\nu=1/3$, and the electron crystal of $N_e=6$ and $N_\phi=21$, respectively. The color bar is truncated at $\bar S(\vb q)=1$ to better present other structures.
	(f) Projected structure factor as a function of $q\equiv\abs{\vb{q}}$ for the three states. The solid purple line is the analytical result for WCs.}
\end{figure*}

An impurity is attractive for $Z<0$ and repulsive for $Z>0$. Nonetheless, attractive impurities are equivalent to repulsive ones in classical systems after combining with electrons. 
An attractive impurity of charge $Z<0$ traps $n$ electrons and forms an impurity of effective charge $Z+n$. 
As long as $Z+n<0$, the impurity continues to trap more electrons until $Z+n\geq 0$. 
Hence, each attractive impurity can trap $-[Z]$ electrons (where $[Z]$ denotes the integer part of $Z$) and becomes a repulsive impurity of charge $Z-[Z]\geq0$. 
Alternatively, we can also reach the same conclusion from an energetic perspective. 
The ground state can be defined through regularizing the Coulomb interaction by replacing $1/r$ with $1/\sqrt{r^2+\delta^2}$ and taking the $\delta\to0$ limit. 
The system is dominated by the binding energy of an impurity trapping $n$ electrons, given by $[n(n-1)/2+Zn]/\delta$. As a result, the total energy is minimized only if $n=-[Z]$. Hence, we only study the repulsive case and take $Z=1$ in classical systems. 

Figures~\ref{Fig:ClassicalWC}(b)-\ref{Fig:ClassicalWC}(d) show how impurities affect the real-space distribution of electrons. 
For a small number of impurities [$N_\text{imp}=70$ in Fig.~\ref{Fig:ClassicalWC}(b)], a single WC is broken into several pieces of hexagonal lattices, which share similar orientations. 
As the number of impurities increases in Fig.~\ref{Fig:ClassicalWC}(c) and~\ref{Fig:ClassicalWC}(d), local order exists throughout the system, but they have different orientations. 
The local order remains pronounced up to $N_{\text{imp}}/N_e\approx 1/2$, suggesting a very strong tendency toward local order, but without true long-range crystalline order. 
To further reveal this transition, we calculate the structure factor defined by
\begin{align}
	S(\vb q)=\frac1{N_e}\sum_{i,j=1}^{N_e}e^{-i\vb q\cdot(\vb r_i-\vb r_j)}-N_e\delta_{\vb q,\vb 0}.
\end{align}
As shown in Fig.~\ref{Fig:ClassicalWC}(e), the structure factor $S(\vb q)$ of a perfect WC exhibits six peaks at the reciprocal lattice vectors, which is the hallmark of long-range hexagonal crystalline order.
For a small number of impurities in Fig.~\ref{Fig:ClassicalWC}(f), the peaks are at the same positions but broadened. 
This is consistent with their real-space distribution, where local WCs have a coherent orientation. 
As the number of impurities increases, the system loses this coherence, and $S(\vb q)$ features a ring centered at the origin, with a faint vestige of the six peaks, as shown in Fig.~\ref{Fig:ClassicalWC}(g) and~\ref{Fig:ClassicalWC}(h). Moreover, the radius of the ring increases because the WC is compressed by the repulsive impurities, squeezing the real-space distance among electrons.

\subsection{Noninteracting electron crystals in the lowest Landau level}

To bridge the gap between classical WC and strongly correlated quantum phases, we study the properties of noninteracting electron crystals created by periodic potentials in the lowest Landau level (LLL). The magnetic flux $\phi$ threaded through a torus must be an integer multiple of the magnetic flux quantum $\phi_0=h/e$, and thus, the length of a square torus is given by $a=\sqrt{2\pi l_B^2 N_\phi}$ where $N_\phi=\phi/\phi_0$, and $l_B=\sqrt{\hbar/(eB)}$ is the magnetic length. If the gap between LLs is much greater than all other energy scales, including interaction, impurity, and temperature, the low-temperature state only resides in the LLL. In particular, we study the ground state of the following noninteracting Hamiltonian,
\begin{align}
	H=-\sum_{i=1}^{N_e}P_{\text{LLL}}\delta(\hat {\vb r}-\vb r_i)P_{\text{LLL}},
\end{align}
where $\hat{\vb r}$ is the position operator, $P_{\text{LLL}}$ is the projection operator of the LLL, and $\vb r_i$'s are placed periodically.

In Fig.~\ref{Fig:Compare}(a), we show the density profile of an electron crystal of $N_e=780$ at a filling factor of $\nu=N_e/N_\phi=1/3$. 
The real-space distribution is akin to that of a classical WC, except that the quantum mechanical electron positions are broadened around each lattice site.  
Each electron is localized and occupies the lowest angular momentum orbital to concentrate itself as much as possible, resulting in a broadening of $\sim l_B$. 
However, even this uncorrelated state exhibits a different structure factor from that of classical WCs. 
In quantum systems, the structure factor is defined as
\begin{align}
	S(\vb q)=\frac1{N_e}\expval{\rho_{-\vb q}\rho_{\vb q}}-N_e\delta_{\vb q,\vb 0},
\end{align}
where $\rho_{\vb q}=\sum_{j=1}^{N_e}e^{i\vb q\cdot \hat{\vb r}_j}$ is the Fourier transform of density operator, and $\hat{\vb r}_j$ is the position operator of this $j$th electron. 
In the context of the FQH effect, it is especially useful to introduce the projected structure factor, which describes the intra-LL excitations. The projected structure factor is defined as
\begin{align}
	\bar S(\vb q)=\frac1{N_e}\expval{\bar \rho_{-\vb q} \bar\rho_{\vb q}}-N_e\delta_{\vb q,\vb 0},
\end{align}
where $\bar \rho_{\vb q}=P_{\text{LLL}}\rho_{\vb q} P_{\text{LLL}}$ is the projected density operator, whose explicit expression is given in Appendix~\ref{Appendix:ProjectedDensityOperator}.  
We also note that for many-body states in the LLL, the two structure factors are related by~\cite{Girvin1986}
\begin{align}
	\bar S(\vb q)=S(\vb q)-1+e^{-q^2l_B^2/2}, 
\end{align}
where $q = \abs{\vb q}$ is the norm of $\vb q$.
$\bar S(\vb q)$ of the electron crystal in Fig.~\ref{Fig:Compare}(d) exhibits the same six peaks as the classical case; however, the peak intensity is significantly reduced (approximately 70 compared to 780 in the classical case) due to broadening. 
In addition to the peaks, there is a ring that is absent in pristine classical WCs. 
To benchmark the quantum WC obtained by exact diagonalization (ED) in the following sections, we also calculate a smaller system of $N_e=6$ and $N_\phi=21$ in Fig.~\ref{Fig:Compare}(b) and~\ref{Fig:Compare}(e), whose $\bar S(\vb q)$ manifests similar peaks and rings. 
The ring is a purely quantum phenomenon, stemming from the ``exchange'' terms in $\bar S(\vb q)$. Because the ground state here is a Slater determinant, $\bar S(\vb q)$ can be decomposed into direct and exchange terms according to Wick's theorem. The direct terms give the classical peaks, while the exchange terms lead to the ring. 
We further prove in Appendix~\ref{Appendix:ProjectedStructureFactorWC} that this ring is analytically given by
\begin{align}
	\bar S(\vb q)=e^{-q^2l_B^2/2}-e^{-q^2l_B^2}
\end{align}
in the $\nu\to0$ limit. In Fig.~\ref{Fig:Compare}(f), we plot the $\bar S(\vb q)$ as a function of $q$. 
For both large and small system sizes, the $\bar S(\vb q)$ perfectly falls on the analytical curve. 
We emphasize that the $\bar S(\vb q)$ of a $\nu=1/3$ FQH state also has a ring, as shown in Fig.~\ref{Fig:Compare}(c), but its origin is distinct. 
In the $\nu=1/3$ FQH state, the peak of $\bar S(\vb q)$ is related to the minimum of the magnetoroton modes~\cite{Girvin1986}. 
Hence, the ring peaks at $l_B\abs{\vb q}\approx 1.4$ in the $\nu=1/3$ FQH state but at $l_B\abs{\vb q}=\sqrt{2\ln 2}\approx 1.17$ in electron crystals, as shown in Fig.~\ref{Fig:Compare}(f). We also note that the peak in the $\nu=1/3$ FQH state has a higher value than that in electron crystals. 

The theoretical results in this section, obtained for well-defined pristine and disordered systems, serve as important benchmarks for understanding and interpreting the exact diagonalization results for the disordered WC and FQHL presented in the next sections, bringing out the interplay of interaction and disorder in the lowest Landau level in the strong-field 2D electron system.

\section{Fractional quantum Hall liquid with random short-range disorder}

In this section, we focus on fractional quantum Hall liquids with random short-range disorder, e.g., random point defects and vacancies. 
As electrons are confined within the LLL, the Hamiltonian is given by
\begin{align}
	H=\frac{1}{2a^2}\sum_{\vb q\neq0}\frac{1}{2\epsilon q}:\bar \rho_{\vb q}^\dag\bar\rho_{\vb q}:+\sum_{\vb q}U_{\vb q}\bar \rho_{\vb q},
\end{align}
where the first term is the Coulomb interaction, the second term is the random disorder, and $:\mathcal O:$ denotes the normal order of operator $\mathcal O$. 
In this work, we set $1/(4\pi\epsilon l_B)=1$ as the energy unit. For the random potential, we take $U_q$ as random complex Gaussian variables satisfying $\expval{U_{\vb q}U_{\vb q'}}_{\text{dis}}=W^2/(2\pi N_\phi)\delta_{\vb q,-\vb q'}$~\cite{Sheng2003}, where $\expval{\cdot}_{\text{dis}}$ denotes average over disorder realizations. 
This choice of random potential corresponds to $\expval{U(\vb r)U(\vb r')}_{\text{dis}}=W^2l_B^2\delta(\vb r-\vb r')$ in the real space, which is the standard Gaussian short-range disorder. 
In the following calculations, we restrict our analysis to a system with $N_\phi = 21$, considering both the $N_e = 7$ case (corresponding to a commensurate FQH filling factor of $\nu = 1/3$) and the $N_e = 6$ case (corresponding to three quasiholes) away from commensurate filling.  
Most of our results from now on focus on exact results for these two situations in the LLL. 

\begin{figure*}[!tbp]
	\center
	\includegraphics[width=0.8\textwidth]{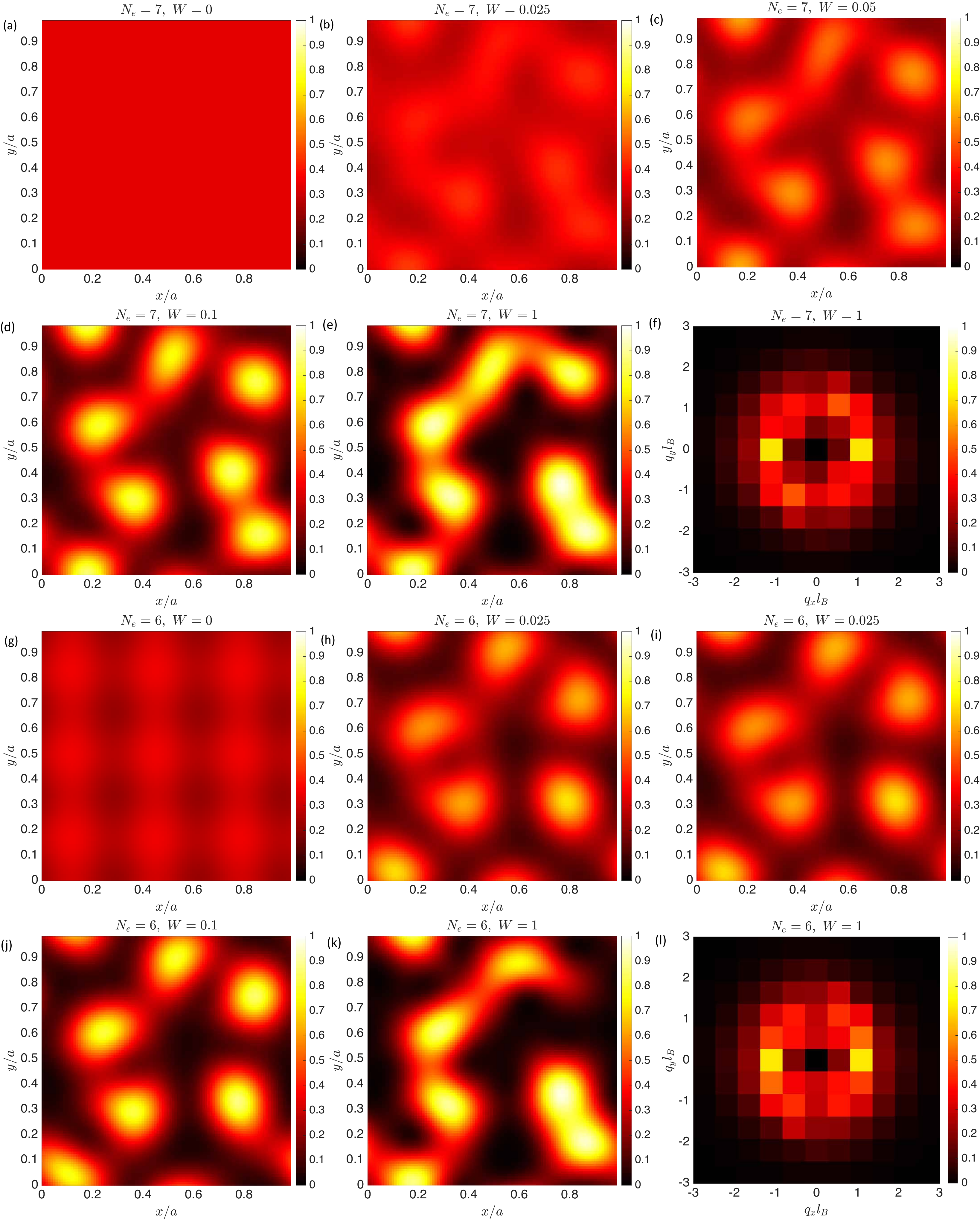}
	\caption{\label{Fig:Random} 
	Calculations of FQH with random disorder in a $N_\phi=21$ system. 
	The top two rows are for $N_e=7$.
	(a)-(d) are the normalized density for increasing disorder strength. 
	(e) and (f) are the normalized density and projected structure factor for $W=1$.
	The bottom two rows are for $N_e=6$. 
	(g)-(j) are the normalized density for increasing disorder strength. 
	(k) and (l) are the normalized density and projected structure factor for $W=1$. 
	Here, we use the same disorder realization for all calculations.
	The projected structure factors for the disorder strengths in (a)-(d) and (g)-(j) are shown in Fig.~\ref{FigSM:Rand0} in the Appendix. 
	}
\end{figure*}

In Fig.~\ref{Fig:Random}, we first calculate the density profile at the exact $1/3$ filling for various disorder strengths using a single disorder realization. 
We note that the experimental low temperature samples typically correspond to single disorder realizations. 
In the absence of disorder, the pristine $1/3$ FQH state has an exactly uniform density protected by the FQH gap, as shown in Fig.~\ref{Fig:Random}(a). 
Weak disorder [Fig.~\ref{Fig:Random}(b)] only slightly perturbs the real-space density, as a result of the incompressibility of the FQH state. 
Nonetheless, for stronger disorder [Fig.~\ref{Fig:Random}(c) and~\ref{Fig:Random}(d)], the electrons become localized, and they are separated from one another because of the Coulomb repulsion. 
In fact, moderate disorder enhances the WC phase since electrons can adjust their positions to lower the energy in the compressible WC phase, but not so much in the incompressible FQH phase. Strong disorder, however, leads to strong localization in both phases.

\begin{figure}[!tbp]
	\center
	\includegraphics[width=0.45\textwidth]{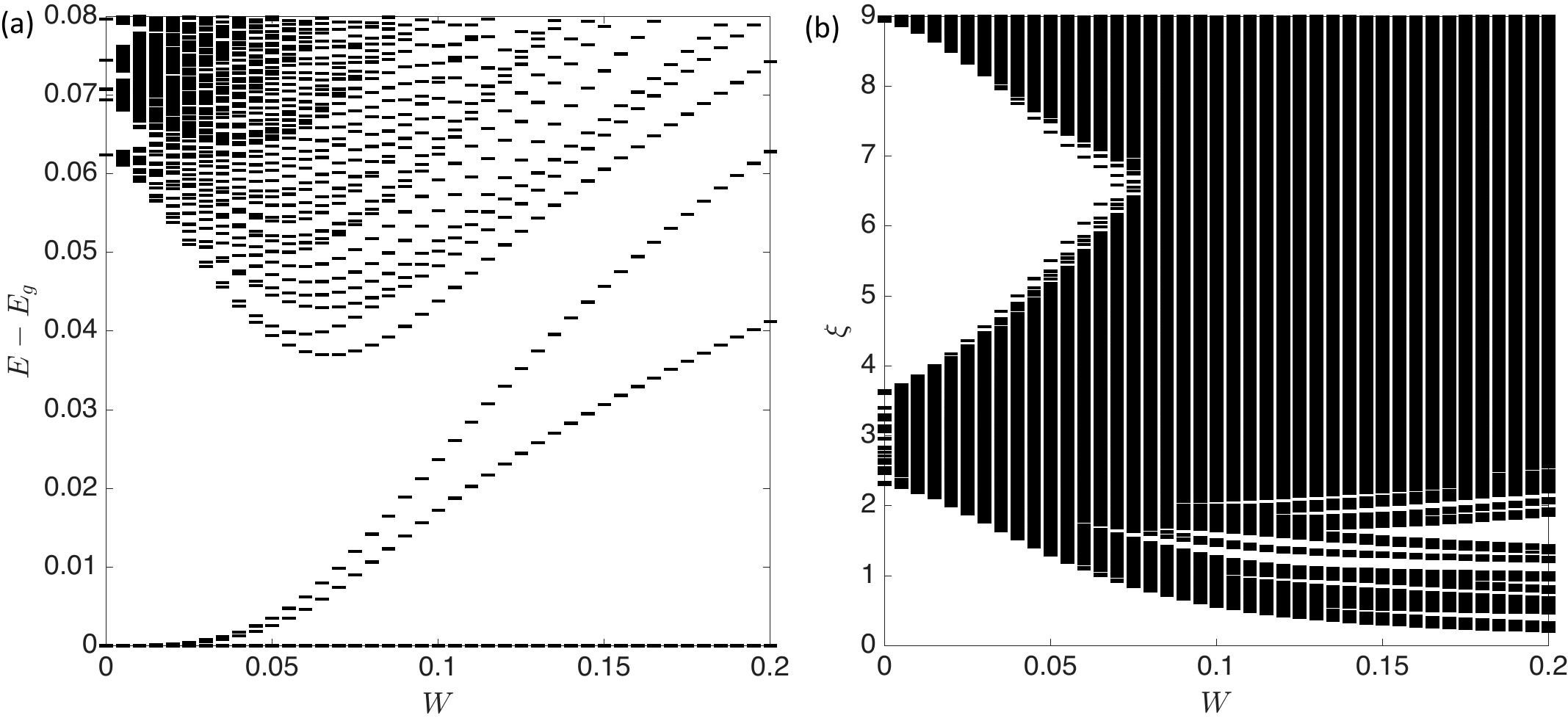}
	\caption{\label{Fig:Random_spectrum} 
	(a) and (b) are the energy spectrum and particle entanglement spectrum of the $N_e=7$ case in a $N_\phi=21$ system with random disorder, respectively. The same disorder realization as in Fig.~\ref{Fig:Random} is used. 
	}
\end{figure}

The energy spectrum in Fig.~\ref{Fig:Random_spectrum}(a) further corroborates the localization transition. 
For $W=0$, the FQH ground state at $1/3$ filling is three-fold degenerate and gapped from excitations, which are the characteristics of topological incompressible states. 
Though the energy gap does not close up to at least $W=0.2$ (note that the approximate pristine FQH gap $\sim 0.06$) in this finite-size system, the drastic increase of the energy splitting beyond $W=0.05$ suggests the destruction of the topological ground state, consistent with the scaling analysis in Ref.~\cite{Sheng2003}. 
It is interesting that the FQHE is destroyed not by the total suppression of the gap, which may happen for much larger $W$, but by the destruction of the topological degeneracy.
So, a trivial gap may survive long after topology is destroyed via the lifting of the topological degeneracy, which happens for a disorder strength ($\sim 0.05$) comparable to the pristine gap ($\sim 0.06$). 

The nature of the FQH ground state is better revealed by the particle entanglement spectrum (PES)~\cite{Sterdyniak2011}. 
For a generic state $\rho$, its PES is defined as the eigenvalue spectrum of
$\xi = -\ln \rho_A$, 
where the reduced density matrix $\rho_A=\Tr_B\rho$ is obtained by partitioning the $N_e$ particles into two subsystems, $A$ and $B$, and tracing out subsystem $B$. 
An FQH state is characterized by an entanglement gap in the PES, below which the number of states equals the number of quasihole excitations, and particularly, the excitations of the $1/3$ Laughlin state follow the generalized Pauli principle. 
In Fig.~\ref{Fig:Random_spectrum}(b), we calculate the PES of the density matrix of the lowest three eigenstates $\rho=\sum_{i=1}^{3}\dyad{e_i}/3$. 
For $W<0.075$, the PES possesses an entanglement gap with 637 states below the gap, in accordance with the generalized Pauli principle, and the gap closes for $W>0.075$, corroborating the transition away from FQHE.

As disorder strength further increases [Fig.~\ref{Fig:Random}(e)], the electrons remain localized but are deformed into an amorphous structure. Compared with Fig.~\ref{Fig:Random}(d), the electrons exhibit no sign of repulsion and form arcs, similar to the ones observed in experiments~\cite{Tsui2024}. 
To summarize, the incompressible FQH state is robust against random disorder up to some finite disorder strength. 
Beyond this transition point, the Coulomb interaction first dominates the system, leading to a localized, short-range-ordered ground state, which could be construed as a disordered `local' crystalline state. 
Eventually, the disorder dominates the system for sufficiently strong disorder, creating a localized, amorphous ground state with no discernible short-range order except perhaps on the scale of the inter-particle separation. 
The plot of $\bar{S}(\vb q)$ in Fig.~\ref{Fig:Random}(f) further corroborates the transition. Specifically, the $\bar{S}(\vb q)$ of the localized state at $W=1$ has a ring structure, similar to that of electron crystals in Fig.~\ref{Fig:Compare}(d) and~\ref{Fig:Compare}(e), but with a much smaller peak value, indicating a lack of clear ordered structure. 

We emphasize that the `transition' being discussed above (and in fact, in all our results) is a crossover since finite-size calculations cannot distinguish between a transition and a crossover, but more generally it is possible, even likely, that the phase changes induced by disorder are all crossover phenomena in this system.

Having discussed the commensurate $\nu=1/3$ case, we now investigate how quasiholes interplay with random disorder by going away from the precise $1/3$ commensurate filling. 
In Fig.~\ref{Fig:Random}(g), we show the density profile for the quasihole case away from the precise commensurate filling in the absence of disorder. 
Unlike Fig.~\ref{Fig:Random}(a), the density here is not uniform because of the interference between quasihole excitations with different momenta. 
Though the common eigenstates of the Hamiltonian and translation operators have uniform density, the ground states are 7-fold degenerate, and we take an arbitrary superposition of them in Fig.~\ref{Fig:Random}(g). 
Therefore, the resulting interference appears as a periodic modulation in real space. 
This is consistent with the general belief that away from the precise odd denominator fillings, the pristine strong field LLL 2D system is in fact a WC in order to minimize the electronic Coulomb energy. 
The abundant quasiholes also make the system more susceptible to disorder and prone to localization. 
As shown in Fig.~\ref{Fig:Random}(h), the system becomes localized at a much smaller $W$ than the exact $\nu=1/3$ case. In the Coulomb-dominant localized regime [Fig.~\ref{Fig:Random}(h)-\ref{Fig:Random}(j)], the density in the presence of quasiholes resembles that in the exact $\nu=1/3$ case, with each electron localized and separated. 
Furthermore, the ground state also becomes amorphous for very strong disorder [see Fig.~\ref{Fig:Random}(k) and \ref{Fig:Random}(l)] as it does in the exact $\nu=1/3$ case.
Hence, the quasihole case also experiences the crossover from local WCs to amorphous structures, similar to the exact $\nu=1/3$ case, except now, away from the commensurate filling, the WC phase is more stable already for low disorder.

Note that we mainly focus on the density profile in the above discussions. 
The accompanying $\bar S(\vb q)$ for the disorder strengths in Fig.~\ref{Fig:Random}(a)-\ref{Fig:Random}(d) and Fig.~\ref{Fig:Random}(g)-\ref{Fig:Random}(j) are shown in Fig.~\ref{FigSM:Rand0} in the Appendix. 
Finally, we also present the results for two more disorder realizations in Fig.~\ref{FigSM:Rand1} and Fig.~\ref{FigSM:Rand2} in the Appendix, which show similar crossovers.

\section{Fractional quantum Hall liquid with charged impurities}

\begin{figure*}[!tbp]
	\center
	\includegraphics[width=0.9\textwidth]{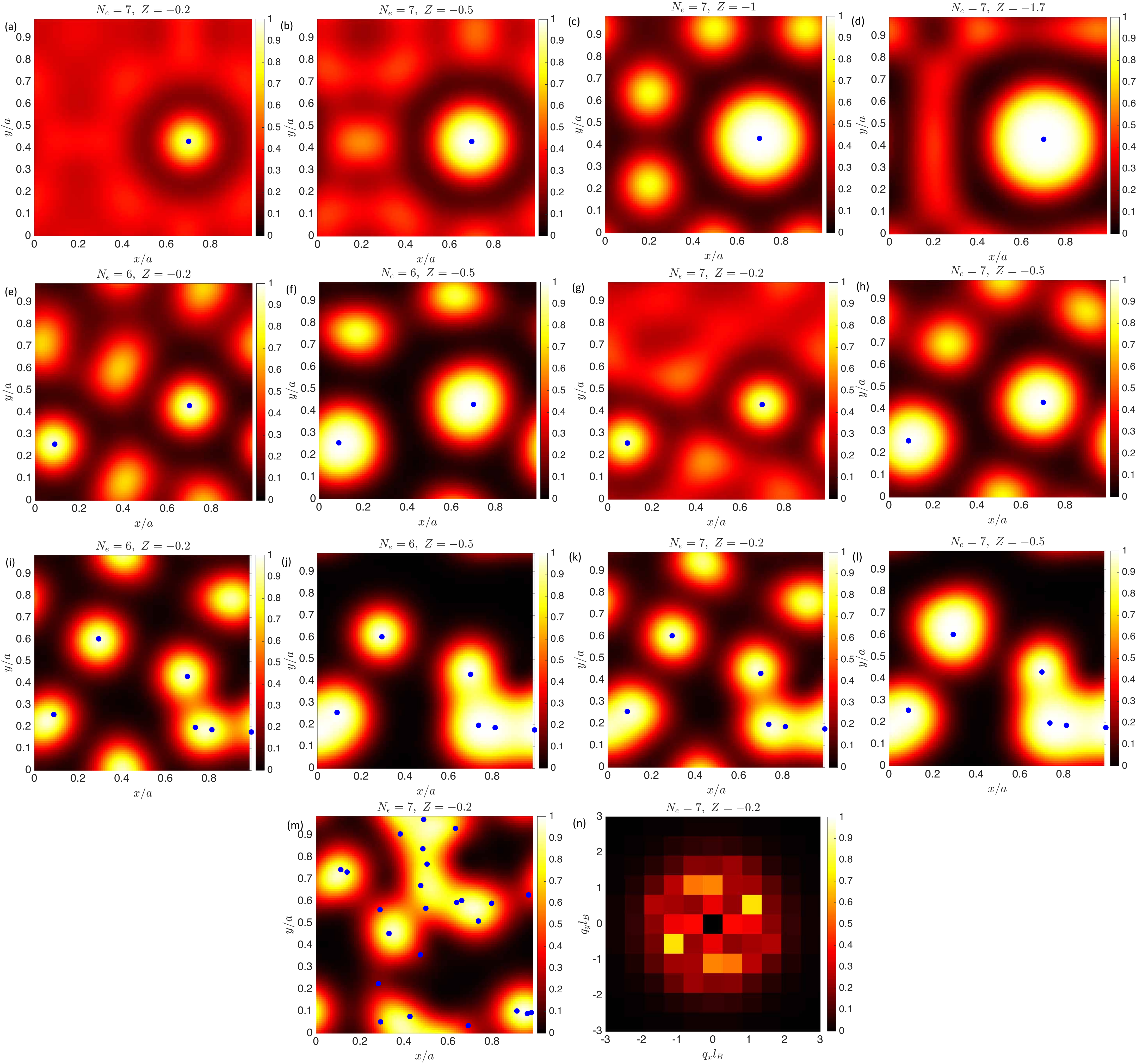}
	\caption{\label{Fig:Coulomb} 
	Normalized density (a-m) and projected structure factor (n) of FQH with charged impurities, with $N_e$ and impurity charge $Z$ labeled on the top of each figure. The blue dots mark the position of impurities. The first row is for one impurity, the second row for two impurities, the third row for six impurities, and the last row for $24$ impurities. 
	The corresponding projected structure factors are shown in Fig.~\ref{FigSM:Coulomb} in the Appendix. 
	}
\end{figure*}

In this section, we study the interplay between FQHL and charged impurities (i.e., long-range disorder). 
We mention that the random quenched unintentional charged impurities in the background are well-established to be the dominant source of disorder in high-quality 2D systems manifesting FQHE~\cite{DasSarma2015,Hwang2008,Ahn2022}.
The Hamiltonian of a system with $N_{\text{imp}}$ impurities is given by
\begin{align}
	H=\frac{1}{2a^2}\sum_{\vb q\neq0}\frac{1}{2\epsilon q}:\bar \rho_{\vb q}^\dag\bar\rho_{\vb q}:+\sum_{\vb q\neq 0}V_{\vb q} \bar\rho_{\vb q},
\end{align}
with 
\begin{align}
	V_{\vb q}=\frac{Z}{2a^2\epsilon q}\sum_{j=1}^{N_{\text{imp}}}e^{-i\vb q\cdot \vb r_j^{\text{imp}}},
\end{align}
where $\vb r_j^{\text{imp}}$ is the position of the $j$th impurity, and $Z$ is the charge of the impurities. 
The positions of the quenched impurities are chosen randomly. References~\cite{Zhang1985,Rezayi1985} first studied the effect of a charged impurity on the incompressible FQH state, and references~\cite{Mostaan2026,Wagner2026,huang2025} have extensively studied the repulsive case in the presence of quasiholes, where impurities pin quasiholes, and the ground state behaves like an incompressible FQH state. Here, we focus on the attractive case.

We start by studying the FQHL with one impurity for the exact $\nu=1/3$ in case of seven electrons in Fig.~\ref{Fig:Coulomb}(a)-\ref{Fig:Coulomb}(d). 
It is known that the FQH gap closes around $Z=-0.4$, and that a magnetoroton mode becomes the ground state for $Z<-0.4$~\cite{Rezayi1985, huang2025}. 
The transition can be seen clearly in Fig.~\ref{Fig:Coulomb}(a) and~\ref{Fig:Coulomb}(b), in which the impurity traps one particle for $Z=-0.2$ but two particles for $Z=-0.5$. 
Notwithstanding, the rest of the electrons show liquid-like behavior until $Z=-1$, where the impurity traps three electrons, with the other four electrons forming a short-range order, as shown in Fig.~\ref{Fig:Coulomb}(c). 
This suggests a strong tendency toward local WC upon perturbations. 
For $Z=-1.7$, the impurity can trap four electrons, and the remaining three electrons form a liquid-like state.
Thus, charged impurities induce considerable short-range order, creating a local effective WC. 
It is notable that charged disorder, because of its long-range correlations, induces substantially more WC-like ordering even at 1/3 filling compared with the corresponding short-range disorder case of the previous section.

Next, we study the effects of multiple impurities on both the exact $1/3$ filling and quasihole cases. 
In Fig.~\ref{Fig:Coulomb}(e)-\ref{Fig:Coulomb}(h), we calculate the density profile for $N_{\text{imp}}=2$. 
For $N_e=6$, the FQHL is localized even at $Z=-0.2$, with six electrons confined in six packets, two of which are at the position of the two impurities, as shown in Fig.~\ref{Fig:Coulomb}(e). 
Its real-space density highly resembles that of the artificial electron crystal in Fig.~\ref{Fig:Compare}(b). 
The similarity between the two states is also confirmed by their $\bar S(\vb q)$ in the Appendix. 
This indicates that the ground state becomes a WC pinned by impurities. 
Finally, increasing the magnitude of $Z$ in Fig.~\ref{Fig:Coulomb}(f), each impurity can trap two electrons, but the system still possesses a locally ordered structure. 
Thus, charged impurities considerably enhance local WC formation, most likely because the compressible nature of the WC helps optimize the Coulomb energy. 

For $N_{\text{imp}}=2$ and $N_e=7$, Fig.~\ref{Fig:Coulomb}(g) demonstrates that a small magnitude of $Z$ cannot destroy the incompressible FQH state, as expected. 
Nonetheless, the FQH state is destroyed for strong disorder at $Z=-0.5$, and the system again becomes an almost amorphous pinned WC, as shown in Fig.~\ref{Fig:Coulomb}(h). 
Moreover, the density profile for $N_e=7$ is qualitatively similar to that for $N_e=6$ in Fig.~\ref{Fig:Coulomb}(f). 
Each impurity traps two electrons, and the two untrapped electrons in Fig.~\ref{Fig:Coulomb}(f) make room for an additional electron to squeeze in, leading to Fig.~\ref{Fig:Coulomb}(h), highlighting the compressible nature of the WC.

Increasing the number of $N_{\text{imp}}$ enhances the tendency toward localization, and particularly, we calculate the density profile for $N_{\text{imp}}=6$ in Fig.~\ref{Fig:Coulomb}(i)-\ref{Fig:Coulomb}(l). 
For $Z=-0.2$, the system exhibits a locally ordered structure for both $N_e=6$ and $N_e=7$ cases in Fig.~\ref{Fig:Coulomb}(i) and~\ref{Fig:Coulomb}(k), which is also corroborated by the structure factor in the Appendix. 
Additionally, for $Z=-0.5$, the system becomes amorphous for both $N_e=6$ and $N_e=7$ cases in Fig.~\ref{Fig:Coulomb}(j) and~\ref{Fig:Coulomb}(l). We also note that as the number of impurities increases, the density profile becomes insensitive to small changes in particle number as the system is now amorphous with hardly any local order. 

\begin{figure*}[!htbp]
	\center
	\includegraphics[width=\textwidth]{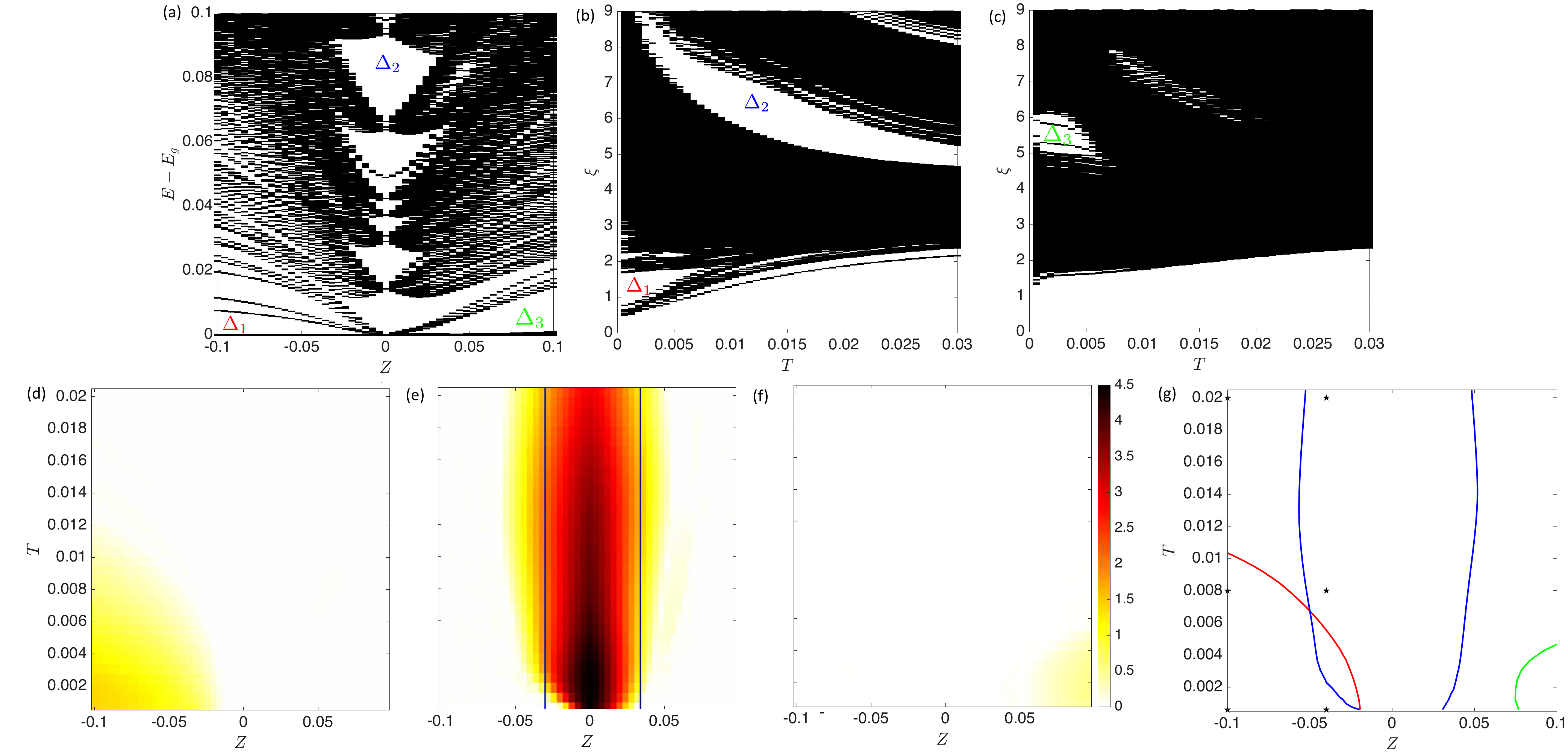}
	\caption{\label{Fig:PhaseDiagram} 
	Calculations in the $N_\phi=21$ system with $N_e=6$ particles and $N_{\text{imp}}=3$ Coulomb impurities, equivalent to $1/3$ FQH state with three quasihole. 
	(a) Energy spectrum as a function of the impurity charge $Z$. There is one state below energy gap 1 (red), corresponding to a single quasihole excitation, 196 states below energy gap 2 (blue), matching the expected number of quasihole excitations, and three states below energy gap 3 (green), corresponding to the FQH state with three quasiholes pinned by three impurities.
	(b) Particle entanglement spectrum at finite temperatures for $Z=-0.04$. There are 637 states below entanglement gap 2, consistent with the generalized Pauli principle for a system of 21 orbitals, and 20 states below entanglement gap 3, consistent with the state counting of a pinned Wigner crystal. (c) Particle entanglement spectrum at finite temperatures for $Z=0.1$. There are 330 states below the entanglement gap 1, consistent with the generalized Pauli principle for a system of 18 orbitals. Here, the particle entanglement spectrum is obtained by retaining $N_a=3$ particles. 
	(d)-(f) are the phase diagrams for the entanglement gaps $\Delta_1$, $\Delta_2$, and $\Delta_3$, respectively. The blue vertical lines in (e) indicate where the energy gap $\Delta_2$ vanishes, and the three figures share the same color bar. (g) Phase boundary determined by the contour lines of the three entanglement gaps equal to $0.3$. 
	The red, blue, and green lines correspond to the entanglement gaps $\Delta_1$, $\Delta_2$, and $\Delta_3$, respectively. The stars show the parameters used in Fig.~\ref{Fig:FiniteT}.
	}
\end{figure*}

It is instructive to consider the large-$N_{\text{imp}}$ limit. 
In this limit, the impurity potential becomes self-averaging, and its behavior is primarily captured by the correlation among $V_{\vb q}$,
\begin{align}
	\expval{V_{\vb q}V_{\vb q'}}_{\text{realization}}=\delta_{\vb q,-\vb q'} [Z/(2a^2\epsilon q)]^2 N_{\text{imp}},
\end{align}
which is similar to the short-range disorder, but with a $1/q^2$ tail.
In particular, there are two meaningful $N_{\text{imp}}\to\infty$ limits: (1) fixing the total charge $ZN_{\text{imp}}$ or (2) fixing $Z$. 
The first limit corresponds to smearing $ZN_{\text{imp}}$ over the whole system, and a uniform charge background is expected. 
Meanwhile, we have the following relation, 
\begin{align*}
\expval{V_{\vb q}V_{\vb q'}}_{\text{realization}}\sim1/N_{\text{imp}}, 
\end{align*}
which vanishes for fixed $ZN_{\text{imp}}$, consistent with the expectation. 
For the second limit, increasing $N_{\text{imp}}$ is analogous to increasing $W$ of random disorder, so we anticipate an amorphous ground state for large $N_{\text{imp}}$. 
In Fig.~\ref{Fig:Coulomb}(m) and~\ref{Fig:Coulomb}(n), we numerically calculate the density profile and projected structure factor for $N_{\text{imp}}=24$ and $Z=-0.2$. 
Although the magnitude of $Z$ is small, the system is distorted into an amorphous structure, qualitatively similar to the strong random disorder case in Fig.~\ref{Fig:Random}(k) and~\ref{Fig:Random}(l).

Apart from the density profile, we also calculate the projected structure factor $\bar S(\vb q)$ for all the cases in Fig.~\ref{Fig:Coulomb}, which are shown in Fig.~\ref{FigSM:Coulomb} in the Appendix.
In sharp contrast to the random-disorder case, $\bar S(\vb q)$ develops pronounced peaks at moderate $\abs{Z}$, reflecting the crystalline order of the impurity-pinned WC evident in the real-space density.
In particular, the peaks for $N_{\text{imp}}=2$ closely resemble those of the artificial electron crystal in Fig.~\ref{Fig:Compare}, corroborating the identification of this regime as a pinned WC.
For large $N_{\text{imp}}$ and $\abs{Z}$, the peaks fade into a ring-like feature, consistent with the crossover to an amorphous structure and qualitatively similar to the strong random-disorder case in Fig.~\ref{Fig:Random}(f).
Qualitatively similar features are observed experimentally~\cite{Tsui2024}.

Hence, the phase diagram in the presence of charged impurities can be summarized as follows.
The incompressible FQH state is resilient to a small number of impurities $N_{\text{imp}}$ with small $\abs{Z}$, while the FQH state with quasiholes is much more vulnerable. 
For intermediate $N_{\text{imp}}$ and $\abs{Z}$, charged impurities lead to local WC structures similar to the random disorder case, but with a much longer correlation length. 
Although the coherent WC pattern spans the entire finite system in Fig.~\ref{Fig:Coulomb}(e), this finite-size observation cannot establish true long-range crystalline order in the thermodynamic limit. 
The Imry-Ma argument suggests that random impurity potentials ultimately fragment the crystal into locally ordered domains rather than preserving a globally coherent WC~\cite{Imry1975}. 
The classical results support this domain picture: for dilute impurities, the crystalline correlation length can be very large, as in Larkin-type pinning physics~\cite{Larkin1970,Larkin1979}. 
Thus, STM images may display apparently long-range WC order over the experimental field of view even when true long-range periodic order is absent asymptotically. 
For large $N_{\text{imp}}$ and $\abs{Z}$, the system becomes amorphous as it does in the presence of random disorder.

\section{Fractional quantum Hall liquid at finite temperatures}
Having established the role of charged impurities at zero temperature, we now turn to finite temperatures, where entropy becomes equally important as energy in shaping the phase diagram. 
In particular, the system may cross over from its zero-temperature ground state to an FQH state with free quasiholes, which carries much higher entropy. 
Ref.~\cite{huang2025} demonstrates that such a crossover can occur for FQH states with pinned quasiholes: at low temperatures, quasiholes are bound to impurities, but above an excitation temperature scale $T_e = \delta E / \delta S$, where $\delta E$ and $\delta S$ are the energy cost and entropy gain of ionization, respectively, they are thermally released. 
A crossover takes place provided $T_e$ lies below the FQH temperature scale $T_{\text{FQH}}$, above which the FQH state itself is thermally suppressed (i.e., $T_\text{FQH}$ larger than the FQH energy gap). 
The same mechanism applies, in principle, to the crossover between Wigner crystals and FQH states with free quasiholes. 
There, impurities trap electrons at zero temperature, reducing the itinerant carrier density and suppressing FQH order. 
For $T>T_e$, however, the trapped electrons are thermally ionized, restoring the itinerant density and enabling FQH state formation for temperatures below the intrinsic FQH incompressible gap. 

\begin{figure*}[t]
	\center
	\includegraphics[width=0.9\textwidth]{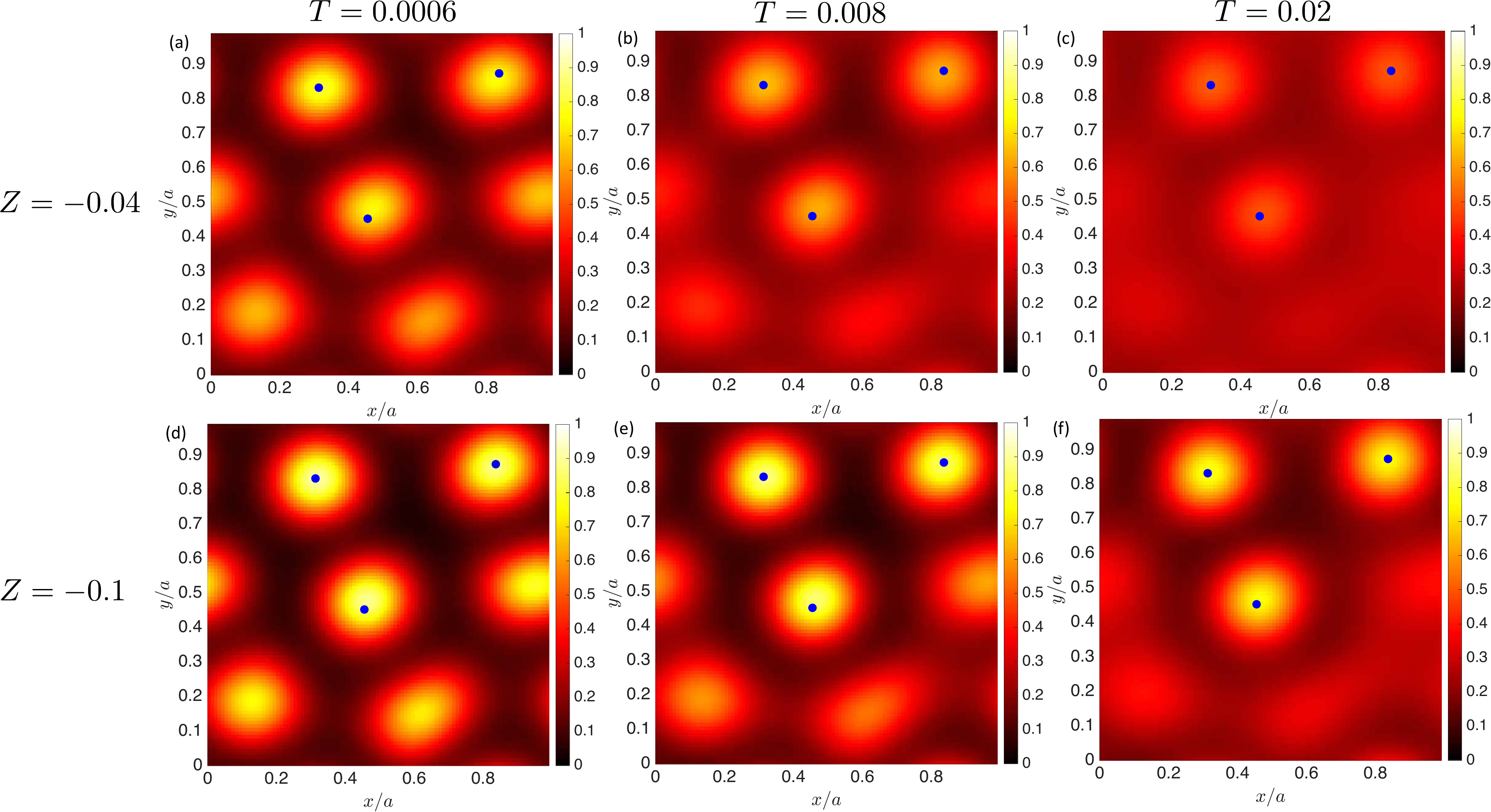}
	\caption{\label{Fig:FiniteT} 
	Normalized density at finite temperatures. 
	The six panels correspond to the six stars in Fig.~\ref{Fig:PhaseDiagram}(g), where we have specified the values of $T$ and $Z$ for each panel. The blue dots mark the position of impurities. 
	}
\end{figure*}

To verify this heuristic argument, we study an FQH system of three quasiholes and three impurities in Fig.~\ref{Fig:PhaseDiagram}. 
We first show the energy spectrum as a function of impurity charge $Z$ in Fig.~\ref{Fig:PhaseDiagram}(a). 
At $Z=0$, the $196$ quasihole excitations, dictated by the generalized Pauli principle, form a low-energy manifold with a finite energy gap $\Delta_2$ above it. 
For the attractive case ($Z<0$), there is a unique gapped ground state below the energy gap $\Delta_1$, and we identify it as a pinned WC in conjunction with the density profile in Fig.~\ref{Fig:FiniteT}(a). 
For the repulsive case ($Z>0$), three nearly degenerate states lie below the energy gap $\Delta_3$, indicating that the ground state is an incompressible FQH state at $\nu=1/3$. 
This occurs because the three impurity orbitals are forbidden from being occupied, leaving an effective system of $N_e=6$ electrons with $N_\phi=18$ flux quanta and an effective filling factor $\nu=1/3$~\cite{huang2025}. 

Further, we study the system at finite temperatures through the PES of the finite-temperature state given by $\rho=e^{-H/T}/\trace[e^{-H/T}]$. 
In Fig.~\ref{Fig:PhaseDiagram}(b), we calculate the PES for $Z=-0.04$. 
At low temperature, there is an entanglement gap $\Delta_1$ above 20 states in the PES, consistent with the state counting of a pinned WC. 
The PES thus further corroborates the WC ground state for the attractive case. 
However, as the temperature increases, the entanglement gap $\Delta_1$ closes and a new entanglement gap $\Delta_2$ emerges with 637 states below it. This is the same entanglement gap that appears at $Z=0$.
The state counting of the entanglement gap $\Delta_2$ also follows the generalized Pauli principle for a system of 21 orbitals, and thus, we conclude that the system crosses over from a WC at low temperatures to a FQH state with free quasiholes at higher temperatures. 
For the repulsive case, we calculate the system at $Z=0.1$ in Fig.~\ref{Fig:PhaseDiagram}(c). 
At low temperatures, the entanglement gap $\Delta_3$ follows the state counting of the generalized Pauli principle for a system of 18 orbitals instead of 21 orbitals, which confirms that the three impurity orbitals are forbidden from being occupied. 
Notwithstanding, no other sizable entanglement gap appears at higher temperatures, unlike the one-impurity scenario~\cite{huang2025} where the pinned quasihole gets ionized. 
The difference results from a larger energy penalty here, which makes $T_e>T_{\text{FQH}}$, so the FQH state is destroyed before ionization.

In Fig.~\ref{Fig:PhaseDiagram}(d)-\ref{Fig:PhaseDiagram}(f), we plot the three entanglement gaps as a function of temperature $T$ and charge $Z$, respectively. 
As shown in Fig.~\ref{Fig:PhaseDiagram}(d), the WC entanglement gap $\Delta_1$ only appears at low temperatures and $Z<-0.02$. 
Figure~\ref{Fig:PhaseDiagram}(e) shows the entanglement gap $\Delta_2$, with blue vertical lines marking the closure of the corresponding energy gap. At zero temperature, the entanglement gap necessarily vanishes whenever the energy gap closes, since entropy plays no role. At finite temperatures, by contrast, the entanglement gap can remain finite even beyond the energy-gap closure, sustained by the large entropy of quasihole states.
In Fig.~\ref{Fig:PhaseDiagram}(f), we show the entanglement gap $\Delta_3$, which appears at low temperatures and $Z>0.07$. 
Combining these results, we show the phase diagram in Fig.~\ref{Fig:PhaseDiagram}(g).

Despite abundant information from PES, it is difficult to explore PES in experiments. 
Thus, we study the density profile at finite temperatures in Fig.~\ref{Fig:FiniteT} to provide concrete evidence for the thermal crossover shown by the PES in Fig.~\ref{Fig:PhaseDiagram}(b). 
At $Z=-0.04$, the electrons form a WC at very low temperatures [Fig.~\ref{Fig:FiniteT}(a)]. 
As the temperature increases to $T=0.008$, the system gradually enters the free-quasihole regime [Fig.~\ref{Fig:PhaseDiagram}(g)]: the pinned electrons are partially released from the impurities, while the unpinned electrons become fully extended and liquid-like [Fig.~\ref{Fig:FiniteT}(b)]. 
Deep in the free quasihole regime ($T = 0.02$) [see Fig.~\ref{Fig:FiniteT}(c)], impurities can no longer pin electrons, and the density becomes fairly uniform throughout the system, supporting our heuristic crossover argument. 
In contrast, the finite-temperature density profile has a distinct behavior at $Z=-0.1$. 
At zero temperature, the density profile shown by Fig.~\ref{Fig:FiniteT}(d) looks identical to that at $Z=-0.04$. 
However, the WC structure remains prominent up to $T=0.008$ as shown in Fig.~\ref{Fig:FiniteT}(e), consistent with the phase diagram obtained from the entanglement gap. 
Moreover, the three electrons are strongly pinned by the impurities up to $T=0.02$ as shown in Fig.~\ref{Fig:FiniteT}(f). 
Meanwhile, the unpinned electrons form a liquid-like state, because the system exits the WC regime as shown in the phase diagram Fig.~\ref{Fig:PhaseDiagram}(g). 
Finally, Fig.~\ref{Fig:FiniteT}(f) shows the melting of the WC itself, with the electrons remaining pinned due to the stronger impurity potential with $Z = -0.1$. 

Our results imply that it is conceivable that, at low enough temperatures, in the presence of disorder arising from random charge impurities, the system may undergo a transition from a low-temperature WC phase to a higher-temperature FQH state under suitable conditions, but if the disorder is too strong, then a localized amorphous solid with only short-range order and no FQHE would exist generically. 

\section{Conclusion}
In summary, we have presented a systematic study of the interplay between disorder and interaction in quantum Hall systems, connecting the physics of Wigner crystals and fractional quantum Hall liquids within a unified framework. 
A common motif emerges across all the systems we studied: as disorder strength increases, the ground state evolves from a coherent ordered phase, be it a crystalline solid or an incompressible topological liquid, through an intermediate regime of local crystalline order, and ultimately into an amorphous state. 
This progression is observed in classical Wigner crystals with charged impurities, in quantum Hall liquids with random short-range disorder, and in quantum Hall liquids with long-range Coulomb impurities, suggesting that it is a robust and generic feature of disordered two-dimensional electron systems in strong magnetic fields. 
The arc-like amorphous structures that emerge in the strong-disorder limit bear a striking resemblance to those observed in recent scanning tunneling microscopy experiments on bilayer graphene~\cite{Tsui2024}, suggesting that disorder is their likely origin. 
We also find that the disorder arising from random quenched charge impurities produces more stable and longer-range local crystalline WC order than the short-range disorder situation, perhaps explaining the generic stability of the local WC phase in the STM experiment for intermediate disorder~\cite{Tsui2024}.

In our theory, we vary the amount of disorder (by changing the disorder strength and/or the number of impurities) in the system, thereby directly explaining the disparate behavior observed across samples with different levels of disorder. 
The dirtier the sample, the less likely it is to find FQH and WC phases, with the strongly insulating, disorder-dominated, localized amorphous phase prevailing, as was the case before the discovery of FQHE.  
But our work also applies to the situation where the transitions among the various LLL phases are induced in the same sample by tuning only the filling factor as in the recent STM experiment~\cite{Tsui2024}. 
In particular, decreasing the 2D electron density reduces the LLL filling factor, but also enhances the disorder, since the ratio $n_e/n_i$ decreases, where $n_e$ and $n_i$ are the 2D electron and impurity densities, respectively. 
Thus, decreasing the filling factor is equivalent to increasing disorder. 
The same is true if the filling factor is reduced by increasing the magnetic field, since the physics depends only on the filling factor in the LLL, independent of how it is tuned (whether by changing density, magnetic field, or both).  
At low filling factors, e.g., $1/5$, the exact diagonalization calculation becomes computationally prohibitive, making us focus on the $1/3$ filling factor, but the physics remains the same at other filling factors, with reduced filling factor automatically implying an increasing disorder in the same sample. 
This is why samples showing excellent $1/3$ FQHE may not manifest the $1/5$ FQHE, and better samples showing $1/5$ FQHE may not manifest the $1/7$ or $1/9$ FQHE, etc. 
We emphasize that decreasing the LLL filling factor always enhances the dimensionless disorder strength generically, and indeed this is always true experimentally, where even the highest quality samples eventually manifest an extremely strong insulating phase with no FQHE at low fillings.

Our findings are consistent with the recent STM experiment~\cite{Tsui2024}: precisely at the $1/3$ filling the FQHL phase is relatively stable, but as soon as one moves away from the precise $1/3$ filling, the system tends to form an ordered compressible WC phase in the presence of disorder, and increasing disorder eventually converts the system into a strongly localized disorder-driven amorphous phase as observed experimentally. 
Thus, disorder invariably destabilizes the FQH phase at commensurate fractional fillings, but stabilizes the WC phase away from the commensurate filling, while strong enough disorder eventually suppresses both FQH and WC phases, leading to an insulating strongly localized amorphous solid phase. 
This behavior is generic for both short- and long-range disorder, although the details are different for the two situations. 
The strong-disorder localized amorphous solid phase has mainly very short-range spatial order and no FQHE. 
We also predict a thermally induced reentrant FQHL due to the competition between energy and entropy at finite temperature. 

We conclude by mentioning that the physics discussed in our work (and observed experimentally) is mostly crossover physics with increasing disorder since Imry-Ma arguments rule out any long-range ordering in the presence of disorder. 
Thus, the emergent WC is always an effective ``local'' WC with a Larkin (or coherence) length decreasing with increasing disorder. 
When this length becomes comparable to the pristine WC lattice constant, the system is simply an amorphous solid, with disorder-induced localization dominating the physics and leading to a strongly insulating localized system. 
Whether this phase is called a ``strongly pinned WC'' or ``an amorphous solid'' is a semantic rather than physical distinction. In this regime, it is not a crystal. 
It has no Goldstone mode, and at long wavelength there is a gap rather than a dispersing phonon mode because the broken symmetry in the presence of impurities is not spontaneous: the Hamiltonian has already lost its translational invariance explicitly because of impurity-induced disorder. 

\emph{Acknowledgement}.— K.H. and X.L. are supported by the Research Grants Council of Hong Kong (Grants No. CityU 11300421, CityU 11304823, CityU 11312825, C7012-21G, and C7015-24G) and City University of Hong Kong (Project No. 9610428). 
S.D.S. is supported by the Laboratory for Physical Sciences through the Condensed Matter Theory Center (CMTC) at the University of Maryland.

\bibliography{FQH_v4.bib}

\begin{thebibliography}{35}%
\makeatletter
\providecommand \@ifxundefined [1]{%
 \@ifx{#1\undefined}
}%
\providecommand \@ifnum [1]{%
 \ifnum #1\expandafter \@firstoftwo
 \else \expandafter \@secondoftwo
 \fi
}%
\providecommand \@ifx [1]{%
 \ifx #1\expandafter \@firstoftwo
 \else \expandafter \@secondoftwo
 \fi
}%
\providecommand \natexlab [1]{#1}%
\providecommand \enquote  [1]{``#1''}%
\providecommand \bibnamefont  [1]{#1}%
\providecommand \bibfnamefont [1]{#1}%
\providecommand \citenamefont [1]{#1}%
\providecommand \href@noop [0]{\@secondoftwo}%
\providecommand \href [0]{\begingroup \@sanitize@url \@href}%
\providecommand \@href[1]{\@@startlink{#1}\@@href}%
\providecommand \@@href[1]{\endgroup#1\@@endlink}%
\providecommand \@sanitize@url [0]{\catcode `\\12\catcode `\$12\catcode
  `\&12\catcode `\#12\catcode `\^12\catcode `\_12\catcode `\%12\relax}%
\providecommand \@@startlink[1]{}%
\providecommand \@@endlink[0]{}%
\providecommand \url  [0]{\begingroup\@sanitize@url \@url }%
\providecommand \@url [1]{\endgroup\@href {#1}{\urlprefix }}%
\providecommand \urlprefix  [0]{URL }%
\providecommand \Eprint [0]{\href }%
\providecommand \doibase [0]{https://doi.org/}%
\providecommand \selectlanguage [0]{\@gobble}%
\providecommand \bibinfo  [0]{\@secondoftwo}%
\providecommand \bibfield  [0]{\@secondoftwo}%
\providecommand \translation [1]{[#1]}%
\providecommand \BibitemOpen [0]{}%
\providecommand \bibitemStop [0]{}%
\providecommand \bibitemNoStop [0]{.\EOS\space}%
\providecommand \EOS [0]{\spacefactor3000\relax}%
\providecommand \BibitemShut  [1]{\csname bibitem#1\endcsname}%
\let\auto@bib@innerbib\@empty
\bibitem [{\citenamefont {Tsui}\ \emph {et~al.}(1982)\citenamefont {Tsui},
  \citenamefont {Stormer},\ and\ \citenamefont {Gossard}}]{Tsui1982}%
  \BibitemOpen
  \bibfield  {author} {\bibinfo {author} {\bibfnamefont {D.~C.}\ \bibnamefont
  {Tsui}}, \bibinfo {author} {\bibfnamefont {H.~L.}\ \bibnamefont {Stormer}},\
  and\ \bibinfo {author} {\bibfnamefont {A.~C.}\ \bibnamefont {Gossard}},\
  }\bibfield  {title} {\bibinfo {title} {Two-dimensional magnetotransport in
  the extreme quantum limit},\ }\href
  {https://doi.org/10.1103/physrevlett.48.1559} {\bibfield  {journal} {\bibinfo
   {journal} {Phys. Rev. Lett.}\ }\textbf {\bibinfo {volume} {48}},\ \bibinfo
  {pages} {1559} (\bibinfo {year} {1982})}\BibitemShut {NoStop}%
\bibitem [{\citenamefont {Stormer}(1999)}]{Stormer1999}%
  \BibitemOpen
  \bibfield  {author} {\bibinfo {author} {\bibfnamefont {H.~L.}\ \bibnamefont
  {Stormer}},\ }\bibfield  {title} {\bibinfo {title} {{The fractional quantum
  Hall effect}},\ }\href {https://doi.org/10.1103/revmodphys.71.875} {\bibfield
   {journal} {\bibinfo  {journal} {Rev. Mod. Phys.}\ }\textbf {\bibinfo
  {volume} {71}},\ \bibinfo {pages} {875} (\bibinfo {year} {1999})}\BibitemShut
  {NoStop}%
\bibitem [{\citenamefont {Prange}\ and\ \citenamefont
  {Girvin}(1990)}]{Prange1990}%
  \BibitemOpen
  \bibinfo {editor} {\bibfnamefont {R.~E.}\ \bibnamefont {Prange}}\ and\
  \bibinfo {editor} {\bibfnamefont {S.~M.}\ \bibnamefont {Girvin}},\ eds.,\
  \href@noop {} {\emph {\bibinfo {title} {{The Quantum Hall Effect}}}},\
  \bibinfo {edition} {2nd}\ ed.,\ Graduate Texts in Contemporary Physics\
  (\bibinfo  {publisher} {Springer-Verlag},\ \bibinfo {year}
  {1990})\BibitemShut {NoStop}%
\bibitem [{\citenamefont {Das~Sarma}\ and\ \citenamefont
  {Pinczuk}(1997)}]{DasSarma1997}%
  \BibitemOpen
  \bibinfo {editor} {\bibfnamefont {S.}~\bibnamefont {Das~Sarma}}\ and\
  \bibinfo {editor} {\bibfnamefont {A.}~\bibnamefont {Pinczuk}},\ eds.,\
  \href@noop {} {\emph {\bibinfo {title} {{Perspectives in Quantum Hall
  Effects: Novel Quantum Liquids in Low-Dimensional Semiconductor
  Structures}}}}\ (\bibinfo  {publisher} {John Wiley \& Sons},\ \bibinfo {year}
  {1997})\BibitemShut {NoStop}%
\bibitem [{\citenamefont {Halperin}\ and\ \citenamefont
  {Jain}(2020)}]{Halperin2020}%
  \BibitemOpen
  \bibinfo {editor} {\bibfnamefont {B.~I.}\ \bibnamefont {Halperin}}\ and\
  \bibinfo {editor} {\bibfnamefont {J.~K.}\ \bibnamefont {Jain}},\ eds.,\ \href
  {https://doi.org/10.1142/11751} {\emph {\bibinfo {title} {{Fractional Quantum
  Hall Effects: New Developments}}}}\ (\bibinfo  {publisher} {World
  Scientific},\ \bibinfo {year} {2020})\BibitemShut {NoStop}%
\bibitem [{\citenamefont {Wigner}(1934)}]{Wigner1934}%
  \BibitemOpen
  \bibfield  {author} {\bibinfo {author} {\bibfnamefont {E.}~\bibnamefont
  {Wigner}},\ }\bibfield  {title} {\bibinfo {title} {{On the Interaction of
  Electrons in Metals}},\ }\href {https://doi.org/10.1103/physrev.46.1002}
  {\bibfield  {journal} {\bibinfo  {journal} {Phys. Rev.}\ }\textbf {\bibinfo
  {volume} {46}},\ \bibinfo {pages} {1002} (\bibinfo {year}
  {1934})}\BibitemShut {NoStop}%
\bibitem [{\citenamefont {Laughlin}(1983)}]{Laughlin1983}%
  \BibitemOpen
  \bibfield  {author} {\bibinfo {author} {\bibfnamefont {R.~B.}\ \bibnamefont
  {Laughlin}},\ }\bibfield  {title} {\bibinfo {title} {Anomalous quantum hall
  effect: An incompressible quantum fluid with fractionally charged
  excitations},\ }\href {https://doi.org/10.1103/physrevlett.50.1395}
  {\bibfield  {journal} {\bibinfo  {journal} {Phys. Rev. Lett.}\ }\textbf
  {\bibinfo {volume} {50}},\ \bibinfo {pages} {1395} (\bibinfo {year}
  {1983})}\BibitemShut {NoStop}%
\bibitem [{\citenamefont {Yoshioka}\ \emph {et~al.}(1983)\citenamefont
  {Yoshioka}, \citenamefont {Halperin},\ and\ \citenamefont
  {Lee}}]{Yoshioka1983}%
  \BibitemOpen
  \bibfield  {author} {\bibinfo {author} {\bibfnamefont {D.}~\bibnamefont
  {Yoshioka}}, \bibinfo {author} {\bibfnamefont {B.~I.}\ \bibnamefont
  {Halperin}},\ and\ \bibinfo {author} {\bibfnamefont {P.~A.}\ \bibnamefont
  {Lee}},\ }\bibfield  {title} {\bibinfo {title} {{Ground State of
  Two-Dimensional Electrons in Strong Magnetic Fields and $\frac{1}{3}$
  Quantized Hall Effect}},\ }\href
  {https://doi.org/10.1103/physrevlett.50.1219} {\bibfield  {journal} {\bibinfo
   {journal} {Phys. Rev. Lett.}\ }\textbf {\bibinfo {volume} {50}},\ \bibinfo
  {pages} {1219} (\bibinfo {year} {1983})}\BibitemShut {NoStop}%
\bibitem [{\citenamefont {Lam}\ and\ \citenamefont {Girvin}(1984)}]{Lam1984}%
  \BibitemOpen
  \bibfield  {author} {\bibinfo {author} {\bibfnamefont {P.~K.}\ \bibnamefont
  {Lam}}\ and\ \bibinfo {author} {\bibfnamefont {S.~M.}\ \bibnamefont
  {Girvin}},\ }\bibfield  {title} {\bibinfo {title} {{Liquid-solid transition
  and the fractional quantum-Hall effect}},\ }\href
  {https://doi.org/10.1103/physrevb.30.473} {\bibfield  {journal} {\bibinfo
  {journal} {Phys. Rev. B}\ }\textbf {\bibinfo {volume} {30}},\ \bibinfo
  {pages} {473} (\bibinfo {year} {1984})}\BibitemShut {NoStop}%
\bibitem [{\citenamefont {Levesque}\ \emph {et~al.}(1984)\citenamefont
  {Levesque}, \citenamefont {Weis},\ and\ \citenamefont
  {MacDonald}}]{Levesque1984}%
  \BibitemOpen
  \bibfield  {author} {\bibinfo {author} {\bibfnamefont {D.}~\bibnamefont
  {Levesque}}, \bibinfo {author} {\bibfnamefont {J.~J.}\ \bibnamefont {Weis}},\
  and\ \bibinfo {author} {\bibfnamefont {A.~H.}\ \bibnamefont {MacDonald}},\
  }\bibfield  {title} {\bibinfo {title} {{Crystallization of the incompressible
  quantum-fluid state of a two-dimensional electron gas in a strong magnetic
  field}},\ }\href {https://doi.org/10.1103/physrevb.30.1056} {\bibfield
  {journal} {\bibinfo  {journal} {Phys. Rev. B}\ }\textbf {\bibinfo {volume}
  {30}},\ \bibinfo {pages} {1056} (\bibinfo {year} {1984})}\BibitemShut
  {NoStop}%
\bibitem [{\citenamefont {Zhu}\ and\ \citenamefont {Louie}(1993)}]{Zhu1993}%
  \BibitemOpen
  \bibfield  {author} {\bibinfo {author} {\bibfnamefont {X.}~\bibnamefont
  {Zhu}}\ and\ \bibinfo {author} {\bibfnamefont {S.~G.}\ \bibnamefont
  {Louie}},\ }\bibfield  {title} {\bibinfo {title} {{Wigner crystallization in
  the fractional quantum Hall regime: A variational quantum Monte Carlo
  study}},\ }\href {https://doi.org/10.1103/physrevlett.70.335} {\bibfield
  {journal} {\bibinfo  {journal} {Phys. Rev. Lett.}\ }\textbf {\bibinfo
  {volume} {70}},\ \bibinfo {pages} {335} (\bibinfo {year} {1993})}\BibitemShut
  {NoStop}%
\bibitem [{\citenamefont {Price}\ \emph {et~al.}(1995)\citenamefont {Price},
  \citenamefont {Zhu}, \citenamefont {Das~Sarma},\ and\ \citenamefont
  {Platzman}}]{Price1995}%
  \BibitemOpen
  \bibfield  {author} {\bibinfo {author} {\bibfnamefont {R.}~\bibnamefont
  {Price}}, \bibinfo {author} {\bibfnamefont {X.}~\bibnamefont {Zhu}}, \bibinfo
  {author} {\bibfnamefont {S.}~\bibnamefont {Das~Sarma}},\ and\ \bibinfo
  {author} {\bibfnamefont {P.~M.}\ \bibnamefont {Platzman}},\ }\bibfield
  {title} {\bibinfo {title} {{Laughlin-liquid--Wigner-solid transition at high
  density in wide quantum wells}},\ }\href
  {https://doi.org/10.1103/physrevb.51.2017} {\bibfield  {journal} {\bibinfo
  {journal} {Phys. Rev. B}\ }\textbf {\bibinfo {volume} {51}},\ \bibinfo
  {pages} {2017} (\bibinfo {year} {1995})}\BibitemShut {NoStop}%
\bibitem [{\citenamefont {Yi}\ and\ \citenamefont {Fertig}(1998)}]{Yi1998}%
  \BibitemOpen
  \bibfield  {author} {\bibinfo {author} {\bibfnamefont {H.}~\bibnamefont
  {Yi}}\ and\ \bibinfo {author} {\bibfnamefont {H.~A.}\ \bibnamefont
  {Fertig}},\ }\bibfield  {title} {\bibinfo {title}
  {{Laughlin-Jastrow-correlated Wigner crystal in a strong magnetic field}},\
  }\href {https://doi.org/10.1103/physrevb.58.4019} {\bibfield  {journal}
  {\bibinfo  {journal} {Phys. Rev. B}\ }\textbf {\bibinfo {volume} {58}},\
  \bibinfo {pages} {4019} (\bibinfo {year} {1998})}\BibitemShut {NoStop}%
\bibitem [{\citenamefont {Yang}\ \emph {et~al.}(2001)\citenamefont {Yang},
  \citenamefont {Haldane},\ and\ \citenamefont {Rezayi}}]{Yang2001}%
  \BibitemOpen
  \bibfield  {author} {\bibinfo {author} {\bibfnamefont {K.}~\bibnamefont
  {Yang}}, \bibinfo {author} {\bibfnamefont {F.~D.~M.}\ \bibnamefont
  {Haldane}},\ and\ \bibinfo {author} {\bibfnamefont {E.~H.}\ \bibnamefont
  {Rezayi}},\ }\bibfield  {title} {\bibinfo {title} {{Wigner crystals in the
  lowest Landau level at low-filling factors}},\ }\href
  {https://doi.org/10.1103/physrevb.64.081301} {\bibfield  {journal} {\bibinfo
  {journal} {Phys. Rev. B}\ }\textbf {\bibinfo {volume} {64}},\ \bibinfo
  {pages} {081301} (\bibinfo {year} {2001})}\BibitemShut {NoStop}%
\bibitem [{\citenamefont {Andrei}\ \emph {et~al.}(1988)\citenamefont {Andrei},
  \citenamefont {Deville}, \citenamefont {Glattli}, \citenamefont {Williams},
  \citenamefont {Paris},\ and\ \citenamefont {Etienne}}]{Andrei1988}%
  \BibitemOpen
  \bibfield  {author} {\bibinfo {author} {\bibfnamefont {E.~Y.}\ \bibnamefont
  {Andrei}}, \bibinfo {author} {\bibfnamefont {G.}~\bibnamefont {Deville}},
  \bibinfo {author} {\bibfnamefont {D.~C.}\ \bibnamefont {Glattli}}, \bibinfo
  {author} {\bibfnamefont {F.~I.~B.}\ \bibnamefont {Williams}}, \bibinfo
  {author} {\bibfnamefont {E.}~\bibnamefont {Paris}},\ and\ \bibinfo {author}
  {\bibfnamefont {B.}~\bibnamefont {Etienne}},\ }\bibfield  {title} {\bibinfo
  {title} {{Observation of a Magnetically Induced Wigner Solid}},\ }\href
  {https://doi.org/10.1103/physrevlett.60.2765} {\bibfield  {journal} {\bibinfo
   {journal} {Phys. Rev. Lett.}\ }\textbf {\bibinfo {volume} {60}},\ \bibinfo
  {pages} {2765} (\bibinfo {year} {1988})}\BibitemShut {NoStop}%
\bibitem [{\citenamefont {Goldman}\ \emph {et~al.}(1990)\citenamefont
  {Goldman}, \citenamefont {Santos}, \citenamefont {Shayegan},\ and\
  \citenamefont {Cunningham}}]{Goldman1990}%
  \BibitemOpen
  \bibfield  {author} {\bibinfo {author} {\bibfnamefont {V.~J.}\ \bibnamefont
  {Goldman}}, \bibinfo {author} {\bibfnamefont {M.}~\bibnamefont {Santos}},
  \bibinfo {author} {\bibfnamefont {M.}~\bibnamefont {Shayegan}},\ and\
  \bibinfo {author} {\bibfnamefont {J.~E.}\ \bibnamefont {Cunningham}},\
  }\bibfield  {title} {\bibinfo {title} {{Evidence for two-dimensional quantum
  Wigner crystal}},\ }\href {https://doi.org/10.1103/physrevlett.65.2189}
  {\bibfield  {journal} {\bibinfo  {journal} {Phys. Rev. Lett.}\ }\textbf
  {\bibinfo {volume} {65}},\ \bibinfo {pages} {2189} (\bibinfo {year}
  {1990})}\BibitemShut {NoStop}%
\bibitem [{\citenamefont {Jiang}\ \emph {et~al.}(1990)\citenamefont {Jiang},
  \citenamefont {Willett}, \citenamefont {Stormer}, \citenamefont {Tsui},
  \citenamefont {Pfeiffer},\ and\ \citenamefont {West}}]{Jiang1990}%
  \BibitemOpen
  \bibfield  {author} {\bibinfo {author} {\bibfnamefont {H.~W.}\ \bibnamefont
  {Jiang}}, \bibinfo {author} {\bibfnamefont {R.~L.}\ \bibnamefont {Willett}},
  \bibinfo {author} {\bibfnamefont {H.~L.}\ \bibnamefont {Stormer}}, \bibinfo
  {author} {\bibfnamefont {D.~C.}\ \bibnamefont {Tsui}}, \bibinfo {author}
  {\bibfnamefont {L.~N.}\ \bibnamefont {Pfeiffer}},\ and\ \bibinfo {author}
  {\bibfnamefont {K.~W.}\ \bibnamefont {West}},\ }\bibfield  {title} {\bibinfo
  {title} {{Quantum liquid versus electron solid around $\nu=1/5$ Landau-level
  filling}},\ }\href {https://doi.org/10.1103/physrevlett.65.633} {\bibfield
  {journal} {\bibinfo  {journal} {Phys. Rev. Lett.}\ }\textbf {\bibinfo
  {volume} {65}},\ \bibinfo {pages} {633} (\bibinfo {year} {1990})}\BibitemShut
  {NoStop}%
\bibitem [{\citenamefont {Ahn}\ and\ \citenamefont
  {Das~Sarma}(2023)}]{Ahn2023}%
  \BibitemOpen
  \bibfield  {author} {\bibinfo {author} {\bibfnamefont {S.}~\bibnamefont
  {Ahn}}\ and\ \bibinfo {author} {\bibfnamefont {S.}~\bibnamefont
  {Das~Sarma}},\ }\bibfield  {title} {\bibinfo {title} {{Density-tuned
  effective metal-insulator transitions in two-dimensional semiconductor
  layers: Anderson localization or Wigner crystallization}},\ }\href
  {https://doi.org/10.1103/physrevb.107.195435} {\bibfield  {journal} {\bibinfo
   {journal} {Phys. Rev. B}\ }\textbf {\bibinfo {volume} {107}},\ \bibinfo
  {pages} {195435} (\bibinfo {year} {2023})}\BibitemShut {NoStop}%
\bibitem [{\citenamefont {Babbar}\ \emph {et~al.}(2026)\citenamefont {Babbar},
  \citenamefont {Li},\ and\ \citenamefont {Das~Sarma}}]{Babbar2026}%
  \BibitemOpen
  \bibfield  {author} {\bibinfo {author} {\bibfnamefont {A.}~\bibnamefont
  {Babbar}}, \bibinfo {author} {\bibfnamefont {Z.-J.}\ \bibnamefont {Li}},\
  and\ \bibinfo {author} {\bibfnamefont {S.}~\bibnamefont {Das~Sarma}},\ }\href
  {https://arxiv.org/abs/2601.03521} {\bibinfo {title} {{Wigner solid or
  Anderson solid -- 2D electrons in strong disorder}}} (\bibinfo {year}
  {2026}),\ \Eprint {https://arxiv.org/abs/2601.03521} {arXiv:2601.03521
  [cond-mat.mes-hall]} \BibitemShut {NoStop}%
\bibitem [{\citenamefont {Huang}\ and\ \citenamefont
  {Das~Sarma}(2026)}]{Huang2026}%
  \BibitemOpen
  \bibfield  {author} {\bibinfo {author} {\bibfnamefont {Y.}~\bibnamefont
  {Huang}}\ and\ \bibinfo {author} {\bibfnamefont {S.}~\bibnamefont
  {Das~Sarma}},\ }\href {https://arxiv.org/abs/2601.09687} {\bibinfo {title}
  {{Disorder-induced strong-field strong-localization in 2D systems}}}
  (\bibinfo {year} {2026}),\ \Eprint {https://arxiv.org/abs/2601.09687}
  {arXiv:2601.09687 [cond-mat.mes-hall]} \BibitemShut {NoStop}%
\bibitem [{\citenamefont {Tsui}\ \emph {et~al.}(2024)\citenamefont {Tsui},
  \citenamefont {He}, \citenamefont {Hu}, \citenamefont {Lake}, \citenamefont
  {Wang}, \citenamefont {Watanabe}, \citenamefont {Taniguchi}, \citenamefont
  {Zaletel},\ and\ \citenamefont {Yazdani}}]{Tsui2024}%
  \BibitemOpen
  \bibfield  {author} {\bibinfo {author} {\bibfnamefont {Y.-C.}\ \bibnamefont
  {Tsui}}, \bibinfo {author} {\bibfnamefont {M.}~\bibnamefont {He}}, \bibinfo
  {author} {\bibfnamefont {Y.}~\bibnamefont {Hu}}, \bibinfo {author}
  {\bibfnamefont {E.}~\bibnamefont {Lake}}, \bibinfo {author} {\bibfnamefont
  {T.}~\bibnamefont {Wang}}, \bibinfo {author} {\bibfnamefont {K.}~\bibnamefont
  {Watanabe}}, \bibinfo {author} {\bibfnamefont {T.}~\bibnamefont {Taniguchi}},
  \bibinfo {author} {\bibfnamefont {M.~P.}\ \bibnamefont {Zaletel}},\ and\
  \bibinfo {author} {\bibfnamefont {A.}~\bibnamefont {Yazdani}},\ }\bibfield
  {title} {\bibinfo {title} {Direct observation of a magnetic-field-induced
  {Wigner} crystal},\ }\href {https://doi.org/10.1038/s41586-024-07212-7}
  {\bibfield  {journal} {\bibinfo  {journal} {Nature}\ }\textbf {\bibinfo
  {volume} {628}},\ \bibinfo {pages} {287} (\bibinfo {year}
  {2024})}\BibitemShut {NoStop}%
\bibitem [{\citenamefont {Sheng}\ \emph {et~al.}(2003)\citenamefont {Sheng},
  \citenamefont {Wan}, \citenamefont {Rezayi}, \citenamefont {Yang},
  \citenamefont {Bhatt},\ and\ \citenamefont {Haldane}}]{Sheng2003}%
  \BibitemOpen
  \bibfield  {author} {\bibinfo {author} {\bibfnamefont {D.~N.}\ \bibnamefont
  {Sheng}}, \bibinfo {author} {\bibfnamefont {X.}~\bibnamefont {Wan}}, \bibinfo
  {author} {\bibfnamefont {E.~H.}\ \bibnamefont {Rezayi}}, \bibinfo {author}
  {\bibfnamefont {K.}~\bibnamefont {Yang}}, \bibinfo {author} {\bibfnamefont
  {R.~N.}\ \bibnamefont {Bhatt}},\ and\ \bibinfo {author} {\bibfnamefont
  {F.~D.~M.}\ \bibnamefont {Haldane}},\ }\bibfield  {title} {\bibinfo {title}
  {{Disorder-Driven Collapse of the Mobility Gap and Transition to an Insulator
  in the Fractional Quantum Hall Effect}},\ }\href
  {https://doi.org/10.1103/physrevlett.90.256802} {\bibfield  {journal}
  {\bibinfo  {journal} {Phys. Rev. Lett.}\ }\textbf {\bibinfo {volume} {90}},\
  \bibinfo {pages} {256802} (\bibinfo {year} {2003})}\BibitemShut {NoStop}%
\bibitem [{\citenamefont {Girvin}\ \emph {et~al.}(1986)\citenamefont {Girvin},
  \citenamefont {MacDonald},\ and\ \citenamefont {Platzman}}]{Girvin1986}%
  \BibitemOpen
  \bibfield  {author} {\bibinfo {author} {\bibfnamefont {S.~M.}\ \bibnamefont
  {Girvin}}, \bibinfo {author} {\bibfnamefont {A.~H.}\ \bibnamefont
  {MacDonald}},\ and\ \bibinfo {author} {\bibfnamefont {P.~M.}\ \bibnamefont
  {Platzman}},\ }\bibfield  {title} {\bibinfo {title} {{Magneto-roton theory of
  collective excitations in the fractional quantum Hall effect}},\ }\href
  {https://doi.org/10.1103/physrevb.33.2481} {\bibfield  {journal} {\bibinfo
  {journal} {Phys. Rev. B}\ }\textbf {\bibinfo {volume} {33}},\ \bibinfo
  {pages} {2481} (\bibinfo {year} {1986})}\BibitemShut {NoStop}%
\bibitem [{\citenamefont {Sterdyniak}\ \emph {et~al.}(2011)\citenamefont
  {Sterdyniak}, \citenamefont {Regnault},\ and\ \citenamefont
  {Bernevig}}]{Sterdyniak2011}%
  \BibitemOpen
  \bibfield  {author} {\bibinfo {author} {\bibfnamefont {A.}~\bibnamefont
  {Sterdyniak}}, \bibinfo {author} {\bibfnamefont {N.}~\bibnamefont
  {Regnault}},\ and\ \bibinfo {author} {\bibfnamefont {B.~A.}\ \bibnamefont
  {Bernevig}},\ }\bibfield  {title} {\bibinfo {title} {{Extracting Excitations
  from Model State Entanglement}},\ }\href
  {https://doi.org/10.1103/physrevlett.106.100405} {\bibfield  {journal}
  {\bibinfo  {journal} {Phys. Rev. Lett.}\ }\textbf {\bibinfo {volume} {106}},\
  \bibinfo {pages} {100405} (\bibinfo {year} {2011})}\BibitemShut {NoStop}%
\bibitem [{\citenamefont {Das~Sarma}\ \emph {et~al.}(2015)\citenamefont
  {Das~Sarma}, \citenamefont {Hwang}, \citenamefont {Kodiyalam}, \citenamefont
  {Pfeiffer},\ and\ \citenamefont {West}}]{DasSarma2015}%
  \BibitemOpen
  \bibfield  {author} {\bibinfo {author} {\bibfnamefont {S.}~\bibnamefont
  {Das~Sarma}}, \bibinfo {author} {\bibfnamefont {E.~H.}\ \bibnamefont
  {Hwang}}, \bibinfo {author} {\bibfnamefont {S.}~\bibnamefont {Kodiyalam}},
  \bibinfo {author} {\bibfnamefont {L.~N.}\ \bibnamefont {Pfeiffer}},\ and\
  \bibinfo {author} {\bibfnamefont {K.~W.}\ \bibnamefont {West}},\ }\bibfield
  {title} {\bibinfo {title} {{Transport in two-dimensional modulation-doped
  semiconductor structures}},\ }\href
  {https://doi.org/10.1103/physrevb.91.205304} {\bibfield  {journal} {\bibinfo
  {journal} {Phys. Rev. B}\ }\textbf {\bibinfo {volume} {91}},\ \bibinfo
  {pages} {205304} (\bibinfo {year} {2015})}\BibitemShut {NoStop}%
\bibitem [{\citenamefont {Hwang}\ and\ \citenamefont
  {Das~Sarma}(2008)}]{Hwang2008}%
  \BibitemOpen
  \bibfield  {author} {\bibinfo {author} {\bibfnamefont {E.~H.}\ \bibnamefont
  {Hwang}}\ and\ \bibinfo {author} {\bibfnamefont {S.}~\bibnamefont
  {Das~Sarma}},\ }\bibfield  {title} {\bibinfo {title} {{Limit to
  two-dimensional mobility in modulation-doped GaAs quantum structures: How to
  achieve a mobility of 100 million}},\ }\href
  {https://doi.org/10.1103/physrevb.77.235437} {\bibfield  {journal} {\bibinfo
  {journal} {Phys. Rev. B}\ }\textbf {\bibinfo {volume} {77}},\ \bibinfo
  {pages} {235437} (\bibinfo {year} {2008})}\BibitemShut {NoStop}%
\bibitem [{\citenamefont {Ahn}\ and\ \citenamefont
  {Das~Sarma}(2022)}]{Ahn2022}%
  \BibitemOpen
  \bibfield  {author} {\bibinfo {author} {\bibfnamefont {S.}~\bibnamefont
  {Ahn}}\ and\ \bibinfo {author} {\bibfnamefont {S.}~\bibnamefont
  {Das~Sarma}},\ }\bibfield  {title} {\bibinfo {title} {{Density-dependent
  two-dimensional optimal mobility in ultra-high-quality semiconductor quantum
  wells}},\ }\href {https://doi.org/10.1103/physrevmaterials.6.014603}
  {\bibfield  {journal} {\bibinfo  {journal} {Phys. Rev. Materials}\ }\textbf
  {\bibinfo {volume} {6}},\ \bibinfo {pages} {014603} (\bibinfo {year}
  {2022})}\BibitemShut {NoStop}%
\bibitem [{\citenamefont {Zhang}\ \emph {et~al.}(1985)\citenamefont {Zhang},
  \citenamefont {Vulovic}, \citenamefont {Guo},\ and\ \citenamefont
  {Das~Sarma}}]{Zhang1985}%
  \BibitemOpen
  \bibfield  {author} {\bibinfo {author} {\bibfnamefont {F.~C.}\ \bibnamefont
  {Zhang}}, \bibinfo {author} {\bibfnamefont {V.~Z.}\ \bibnamefont {Vulovic}},
  \bibinfo {author} {\bibfnamefont {Y.}~\bibnamefont {Guo}},\ and\ \bibinfo
  {author} {\bibfnamefont {S.}~\bibnamefont {Das~Sarma}},\ }\bibfield  {title}
  {\bibinfo {title} {{Effect of a charged impurity on the fractional quantum
  Hall effect: Exact numerical treatment of finite systems}},\ }\href
  {https://doi.org/10.1103/physrevb.32.6920} {\bibfield  {journal} {\bibinfo
  {journal} {Phys. Rev. B}\ }\textbf {\bibinfo {volume} {32}},\ \bibinfo
  {pages} {6920} (\bibinfo {year} {1985})}\BibitemShut {NoStop}%
\bibitem [{\citenamefont {Rezayi}\ and\ \citenamefont
  {Haldane}(1985)}]{Rezayi1985}%
  \BibitemOpen
  \bibfield  {author} {\bibinfo {author} {\bibfnamefont {E.~H.}\ \bibnamefont
  {Rezayi}}\ and\ \bibinfo {author} {\bibfnamefont {F.~D.~M.}\ \bibnamefont
  {Haldane}},\ }\bibfield  {title} {\bibinfo {title} {{Incompressible states of
  the fractionally quantized Hall effect in the presence of impurities: A
  finite-size study}},\ }\href {https://doi.org/10.1103/physrevb.32.6924}
  {\bibfield  {journal} {\bibinfo  {journal} {Phys. Rev. B}\ }\textbf {\bibinfo
  {volume} {32}},\ \bibinfo {pages} {6924} (\bibinfo {year}
  {1985})}\BibitemShut {NoStop}%
\bibitem [{\citenamefont {Mostaan}\ \emph {et~al.}(2026)\citenamefont
  {Mostaan}, \citenamefont {Goldman}, \citenamefont {İmamoğlu},\ and\
  \citenamefont {Grusdt}}]{Mostaan2026}%
  \BibitemOpen
  \bibfield  {author} {\bibinfo {author} {\bibfnamefont {N.}~\bibnamefont
  {Mostaan}}, \bibinfo {author} {\bibfnamefont {N.}~\bibnamefont {Goldman}},
  \bibinfo {author} {\bibfnamefont {A.}~\bibnamefont {İmamoğlu}},\ and\
  \bibinfo {author} {\bibfnamefont {F.}~\bibnamefont {Grusdt}},\ }\bibfield
  {title} {\bibinfo {title} {{Anyon-Trions in Atomically Thin Semiconductor
  Heterostructures}},\ }\href {https://doi.org/10.1103/hxmb-pn4z} {\bibfield
  {journal} {\bibinfo  {journal} {PRX Quantum}\ }\textbf {\bibinfo {volume}
  {7}},\ \bibinfo {pages} {010325} (\bibinfo {year} {2026})}\BibitemShut
  {NoStop}%
\bibitem [{\citenamefont {Wagner}\ and\ \citenamefont
  {Neupert}(2026)}]{Wagner2026}%
  \BibitemOpen
  \bibfield  {author} {\bibinfo {author} {\bibfnamefont {G.}~\bibnamefont
  {Wagner}}\ and\ \bibinfo {author} {\bibfnamefont {T.}~\bibnamefont
  {Neupert}},\ }\bibfield  {title} {\bibinfo {title} {{Sensing the binding and
  unbinding of anyons at impurities}},\ }\href
  {https://doi.org/10.1103/cfm5-wgz1} {\bibfield  {journal} {\bibinfo
  {journal} {Phys. Rev. Research}\ }\textbf {\bibinfo {volume} {8}},\ \bibinfo
  {pages} {013263} (\bibinfo {year} {2026})}\BibitemShut {NoStop}%
\bibitem [{\citenamefont {Huang}\ \emph {et~al.}(2025)\citenamefont {Huang},
  \citenamefont {Das~Sarma},\ and\ \citenamefont {Li}}]{huang2025}%
  \BibitemOpen
  \bibfield  {author} {\bibinfo {author} {\bibfnamefont {K.}~\bibnamefont
  {Huang}}, \bibinfo {author} {\bibfnamefont {S.}~\bibnamefont {Das~Sarma}},\
  and\ \bibinfo {author} {\bibfnamefont {X.}~\bibnamefont {Li}},\ }\href
  {https://arxiv.org/abs/2512.07769} {\bibinfo {title} {{Thermal ionization of
  impurity-bound quasiholes in the fractional quantum Hall effect}}} (\bibinfo
  {year} {2025}),\ \Eprint {https://arxiv.org/abs/2512.07769} {arXiv:2512.07769
  [cond-mat.mes-hall]} \BibitemShut {NoStop}%
\bibitem [{\citenamefont {Imry}\ and\ \citenamefont {Ma}(1975)}]{Imry1975}%
  \BibitemOpen
  \bibfield  {author} {\bibinfo {author} {\bibfnamefont {Y.}~\bibnamefont
  {Imry}}\ and\ \bibinfo {author} {\bibfnamefont {S.-K.}\ \bibnamefont {Ma}},\
  }\bibfield  {title} {\bibinfo {title} {{Random-Field Instability of the
  Ordered State of Continuous Symmetry}},\ }\href
  {https://doi.org/10.1103/physrevlett.35.1399} {\bibfield  {journal} {\bibinfo
   {journal} {Phys. Rev. Lett.}\ }\textbf {\bibinfo {volume} {35}},\ \bibinfo
  {pages} {1399} (\bibinfo {year} {1975})}\BibitemShut {NoStop}%
\bibitem [{\citenamefont {Larkin}(1970)}]{Larkin1970}%
  \BibitemOpen
  \bibfield  {author} {\bibinfo {author} {\bibfnamefont {A.~I.}\ \bibnamefont
  {Larkin}},\ }\bibfield  {title} {\bibinfo {title} {{Effect of inhomogeneities
  on the structure of the mixed state of superconductors}},\ }\href@noop {}
  {\bibfield  {journal} {\bibinfo  {journal} {Sov. Phys. JETP}\ }\textbf
  {\bibinfo {volume} {31}},\ \bibinfo {pages} {784} (\bibinfo {year}
  {1970})}\BibitemShut {NoStop}%
\bibitem [{\citenamefont {Larkin}\ and\ \citenamefont
  {Ovchinnikov}(1979)}]{Larkin1979}%
  \BibitemOpen
  \bibfield  {author} {\bibinfo {author} {\bibfnamefont {A.~I.}\ \bibnamefont
  {Larkin}}\ and\ \bibinfo {author} {\bibfnamefont {Y.~N.}\ \bibnamefont
  {Ovchinnikov}},\ }\bibfield  {title} {\bibinfo {title} {{Pinning in type II
  superconductors}},\ }\href {https://doi.org/10.1007/bf00117160} {\bibfield
  {journal} {\bibinfo  {journal} {J. Low Temp. Phys.}\ }\textbf {\bibinfo
  {volume} {34}},\ \bibinfo {pages} {409} (\bibinfo {year} {1979})}\BibitemShut
  {NoStop}%
\end{thebibliography}%

\onecolumngrid
\appendix 

\section{Projected density operator in second quantization language \label{Appendix:ProjectedDensityOperator}}

In this section, we give the projected density operator in terms of creation and annihilation operators. 
Let us first consider LLs on a plane.
The canonical momenta satisfy $[\pi_a,\pi_b]=i\epsilon_{ab}\hbar^2/l_B^2$, where $l_B=\sqrt{\hbar/(eB)}$ is the magnetic length. The LLs can be constructed by a pair of bosonic creation and annihilation operators defined by $a=(\pi_x+i\pi_y)/\sqrt{2}$. Moreover, we consider the guiding center momenta $\vb*Q=\vb*\pi-\bar{\vb*r}$, where we define the Hodge dual $\bar{x}_a=\epsilon_{ab} x_b$. 
The guiding center momenta satisfy the following commutation relations:
\begin{align}
	[r_a,Q_b]=i\delta_{ab},\ \ [\pi_a,Q_b]=0,\ \ [Q_a,Q_b]&=-i\epsilon_{ab}.
\end{align}
The first commutation relation indicates that the magnetic translation $T(\vb* x)=e^{i\vb* x\vdot \vb* Q}$ translates the position operator as the ordinary translation operator. 
The second commutation implies that $T(\vb* x)$ commutes with the creation and annihilation operators, resolving the degeneracy of the LLs. 
However, it is not always possible to construct the common eigenstates of two magnetic translations in the LLL because of the third commutation relation. 
Particularly, we have
\begin{align}
	T(\vb* R_1)T(\vb* R_2)=T(\vb*R_2)T(\vb* R_1)e^{i\vb* R_1 \cross \vb* R_2},
\end{align}
where we define $\vb* x \cross \vb* y=\epsilon_{ab} x_ay_b$. 
There exists a set of common eigenstates in the LLL if $\vb* R_1 \cross \vb* R_2=2\pi l_B^2$, and the momentum of a common eigenstate is defined as
\begin{align}
	T(\vb* R_1)\ket{\psi}=e^{i\vb*k\vdot\vb*R_1}\ket{\psi},\quad T(\vb* R_2)\ket{\psi}=e^{i\vb*k\vdot\vb*R_2}\ket{\psi}.
\end{align}
Moreover, $T(\vb* x)$ also changes the momentum of a state by
\begin{align}
	T(\vb* R_\alpha)T(\vb* x)\ket{\psi}
	=e^{i\vb* R_\alpha \cross \vb* x}T(\vb* x)T(\vb* R_\alpha)\ket{\psi}
	=e^{i(\vb*k+\bar{\vb*x})\vdot\vb*R_\alpha}T(\vb* x)\ket{\psi}.
\end{align}
Consequently, all states in the LLL can be generated by the states with zero momentum through
\begin{align}\label{Eq:Basis}
	\ket{\vb*k,\lambda}=T(-\bar{\vb*k})\ket{\vb*0,\lambda},
\end{align}
where $\lambda$ labels the basis states of an irreducible representation of the little group that does not change the momentum of a state. $T(-\bar{\vb*q})$ does not change the momentum if and only if $\vb*q=m\vb*G_1+n \vb*G_2$, where $m,n$ are two integers, and $\vb G_1=\bar{\vb*R}_2/l_B^2$ and $\vb G_2=-\bar{\vb*R}_1/l_B^2$ are the two reciprocal vectors of $\vb*R_1$ and $\vb*R_2$. Thus, the little group is an Abelian group, given by $\{T(m\vb*R_1+n\vb*R_2):\ m,n\in\mathbb{Z}\}$, and the irreducible representation is one-dimensional. Henceforth, we omit the $\lambda$ index in Eq.~\eqref{Eq:Basis}.

The Hilbert space of the LLL on a torus spanned by $\vb* L_1$ and $\vb* L_2$ is the subspace that satisfies $T(\vb*L_1)\ket{\psi}=T(\vb*L_2)\ket{\psi}=\ket{\psi}$ with $\vb* L_1 \cross \vb* L_2=2\pi N_\phi l_B^2$. We have the freedom to choose $\vb*R_1$ and $\vb*R_2$, and a convenient choice is $\vb*R_1=\vb*L_1$ and $\vb*R_2=\vb*L_2/N_\phi$. The periodic condition translates to the constraint that $\vb*k$ must be integer multiples of $\vb*G_2/N_\phi$, and therefore, the basis states can be denoted by $\ket{n}:=\ket{n \vb*G_2/N_\phi}$ for all $n\in\mathbb{Z}$ with periodicity $\ket{n+N_\phi}=\ket{n}$. 
As a result, the projected density operator is nonvanishing $\vb*q=m\vb*G_2/N_\phi+m'\vb*G_1$ with integers $m,m'$. 
Its explicit expression is given by
\begin{align}
	\bar\rho_{\vb q}=\sum_{n}\mel{n+m}{e^{i\vb*q\vdot\vb*r}}{n}c^\dag_{n+m}c_{n}
\end{align}
with
\begin{align}
	\mel{n+m}{e^{i\vb*q\vdot\vb*r}}{n}
	=\mel{n+m}{e^{i\vb*q\vdot(\bar{\vb*Q}-\bar{\vb*\pi} ) }}{n}
	=e^{i\pi m'(2n+m)/N_\phi}e^{-q^2 l_B^2/4}.
\end{align}

\section{Projected structure factor for Wigner crystals at low densities
\label{Appendix:ProjectedStructureFactorWC}
}

In this section, we present the analytical expression for the projected structure factor $\bar S(\vb q)$ of Wigner crystals at low densities. At very low densities, the electrons reside in the lowest Landau level and form a Wigner crystal (WC), which can be well approximated by a Slater determinant. 
As the projected structure factor is related to the unprojected one through $\bar S(\vb q)=S(\vb q)-1+e^{-q^2 l_B^2/2}$, we choose to first calculate the unprojected structure factor in the real space. Utilizing Wick's theorem, we obtain
\begin{align}
	S(r,r')&=\frac1N\expval{\psi^\dag(r)\psi(r)\psi^\dag(r')\psi(r')} \notag\\
	&=\frac1N\expval{\psi^\dag(r)\psi(r)}\expval{\psi^\dag(r')\psi(r')}
	+\frac1N\expval{\psi^\dag(r)\psi(r')}\expval{\psi(r)\psi^\dag(r')},
\end{align}
where $\psi(r)$ and $\psi^\dag(r)$ are the fermionic field operators. The first (direct) term is identical to the classical structure factor, while the second (exchange) term is absent in classical systems. Using the one-body density matrix $\expval{\psi^\dag(r)\psi(r')}=\sum_i\phi_i^*(r)\phi_i(r')$ and the anticommutator identity $\expval{\psi(r)\psi^\dag(r')}=\delta(r-r')-\sum_j\phi_j^*(r')\phi_j(r)$, where $\phi_i(r)$ denotes the single-particle wave function of the $i$th occupied orbital, the exchange term can be written as
\begin{align}
	S_{\text{ex}}(r,r')&=\frac{1}{N}\delta(r-r')\expval{n(r)}
	-\frac1N\sum_{i,j}\phi_i^*(r)\phi_i(r')\phi_j^*(r')\phi_j(r),
\end{align}
where $\expval{n(r)}=\sum_{i}\abs{\phi_i(r)}^2$ is the density. At very low densities, the wave functions do not overlap with each other, so they satisfy $\phi_i^*(r)\phi_j(r')=\delta_{ij}\phi_i^*(r)\phi_j(r')$. Moreover, as an individual electron tends to localize itself as much as possible, its wave function is approximately described by the LLL eigenstate with the lowest angular momentum, that is $\phi_i(r)\propto e^{-(r-r_i)^2/(4l_B^2)}$, where $r_i$ is the localization center. Thus, we have
\begin{align}
	S_{\text{ex}}(r,r')=\frac{1}{N}\delta(r-r')\expval{n(r)}-\frac1N\sum_{i}\abs{\phi_i(r)}^2\abs{\phi_i(r')}^2,
\end{align}
whose Fourier transform is given by
\begin{align}
	S_{\text{ex}}(\vb q)&=\int\dd{r}\dd{r'}e^{i\vb q\vdot (r'-r)}S_{\text{ex}}(r,r')=1-e^{-q^2l_B^2}.
\end{align}
Following this, we obtain $\bar S_{\text{ex}}(\vb q)=e^{-q^2l_B^2/2}-e^{-q^2l_B^2}$.

\begin{figure*}[!tbp]
	\center
	\includegraphics[width=0.85\textwidth]{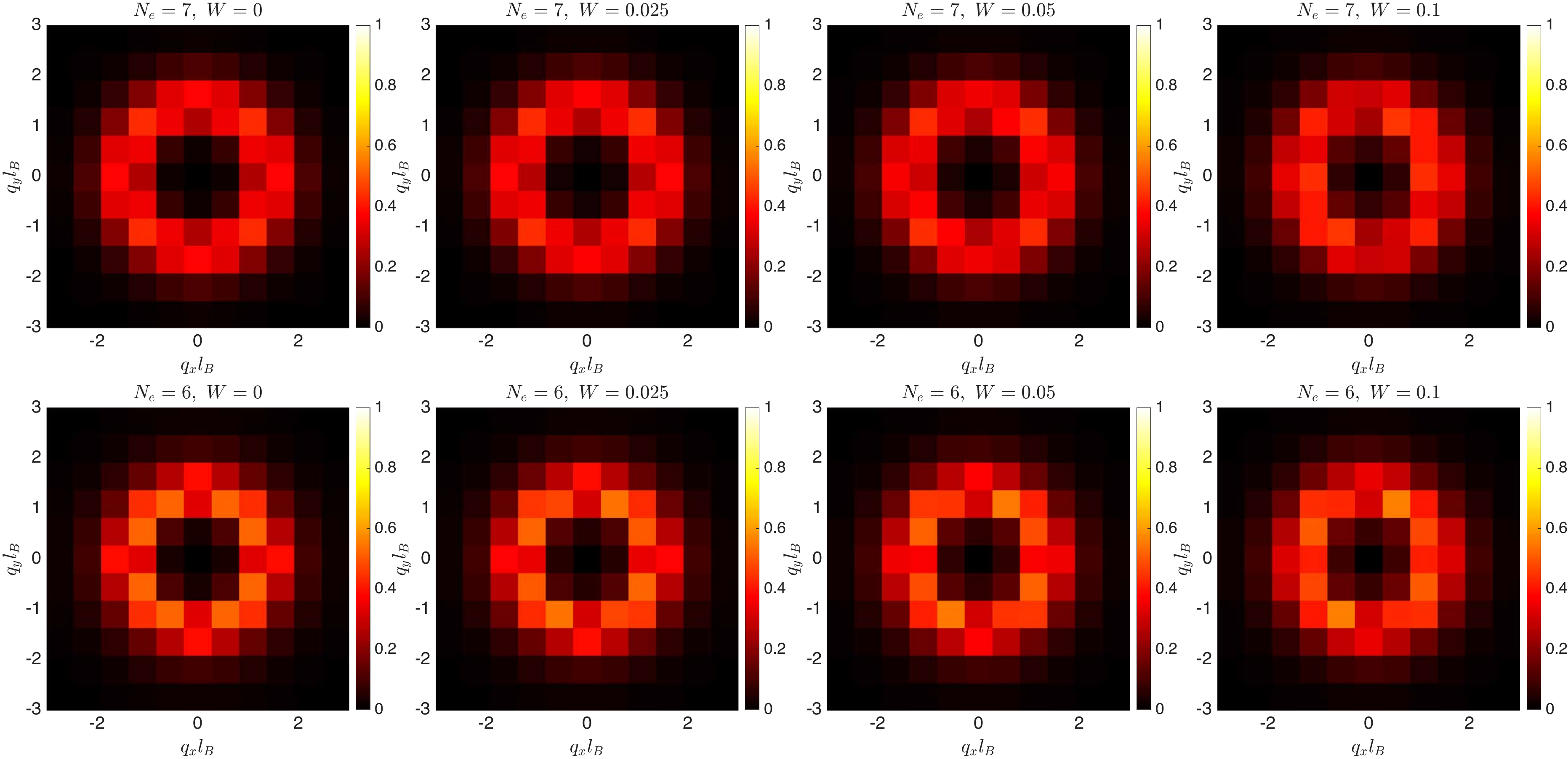}
	\caption{\label{FigSM:Rand0} 
	Projected structure factor for the FQHL at zero temperature with the disorder realization in the main text. Here, we take $N_\phi=21$. 
	These results accompany the density profiles in Fig.~\ref{Fig:Random} in the main text, showing the evolution of the projected structure factor as the disorder strength increases. 
	}
\end{figure*}

\section{Additional results for fractional quantum Hall liquid with random disorder
\label{Appendix:ProjectedStructureFactorRandomDisorder}
}

In this section, we first present the projected structure factor for various disorder strengths in Fig.~\ref{FigSM:Rand0}, using the disorder realization in Fig.~\ref{Fig:Random} in the main text.  
As the disorder strength increases, $\bar S(\vb* q)$ does not have qualitative changes, despite the localization transition indicated by the real-space density. 
The absence of peaks in $\bar S(\vb* q)$ further validates the absence of a local ordered structure.

\begin{figure*}[!tbp]
	\center
	\includegraphics[width=0.85\textwidth]{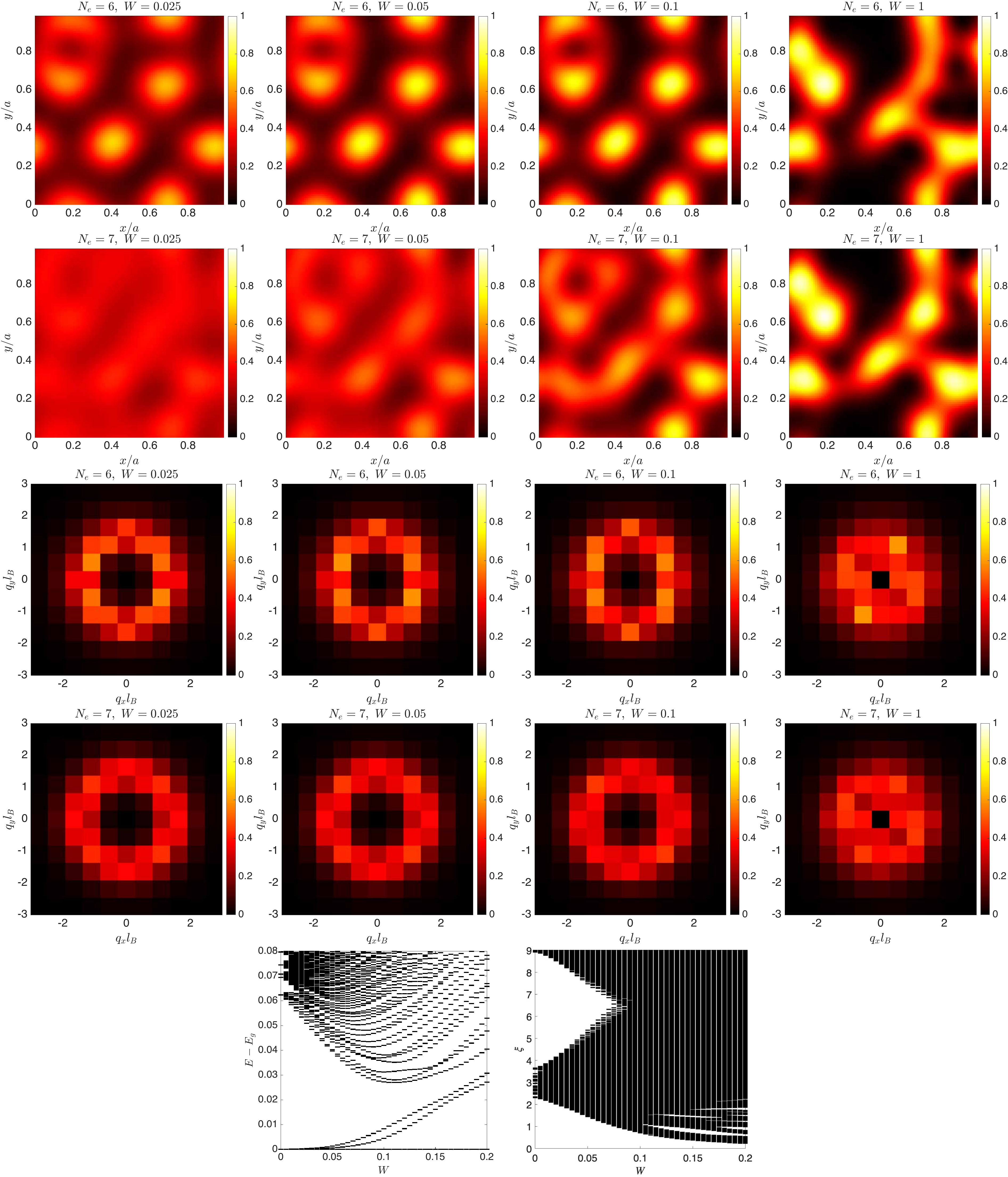}
	\caption{\label{FigSM:Rand1} 
	Density profile and projected structure factor for the FQHL at zero temperature using a second random disorder realization. The first and second rows are the density profile, and the third and fourth rows are the projected structure factor. 
	The last row is the energy spectrum and particle entanglement spectrum. 
	Here, we take $N_\phi=21$. 
	}
\end{figure*}

\begin{figure*}[!tbp]
	\center
	\includegraphics[width=0.85\textwidth]{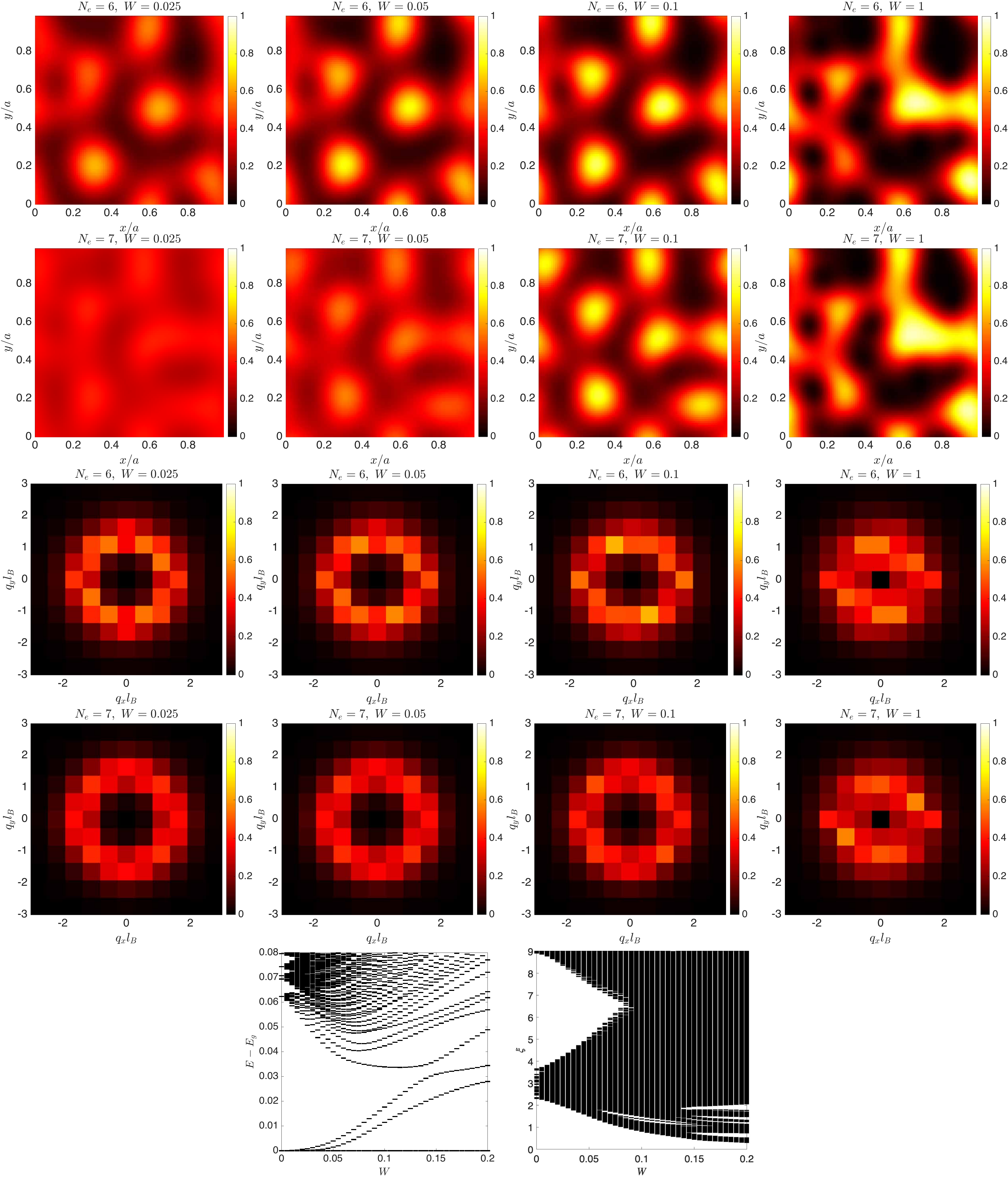}
	\caption{\label{FigSM:Rand2} 
	Density profile and projected structure factor for the FQHL at zero temperature using a third random disorder realization. The first and second rows are the density profile, and the third and fourth rows are the projected structure factor. 
	The last row is the energy spectrum and particle entanglement spectrum. 
	Here, we take $N_\phi=21$.
	}
\end{figure*}

Furthermore, we present the density profile and projected structure factor for two more disorder realizations in Fig.~\ref{FigSM:Rand1} and Fig.~\ref{FigSM:Rand2}. 
The results exhibit physics similar to those in the main text, demonstrating the generality of the conclusion. 
We also show the energy spectrum and the particle entanglement spectrum (PES) for the two disorder realizations in each figure. 
Similar to the results in the main text, the FQH gap does not close, but the degeneracy of the lowest three states is lifted, suggesting a phase transition. The PES also demonstrates the same transition.

\section{Projected structure factor for fractional quantum Hall liquid with charged impurities}

\begin{figure*}[!tbp]
	\center
	\includegraphics[width=0.85\textwidth]{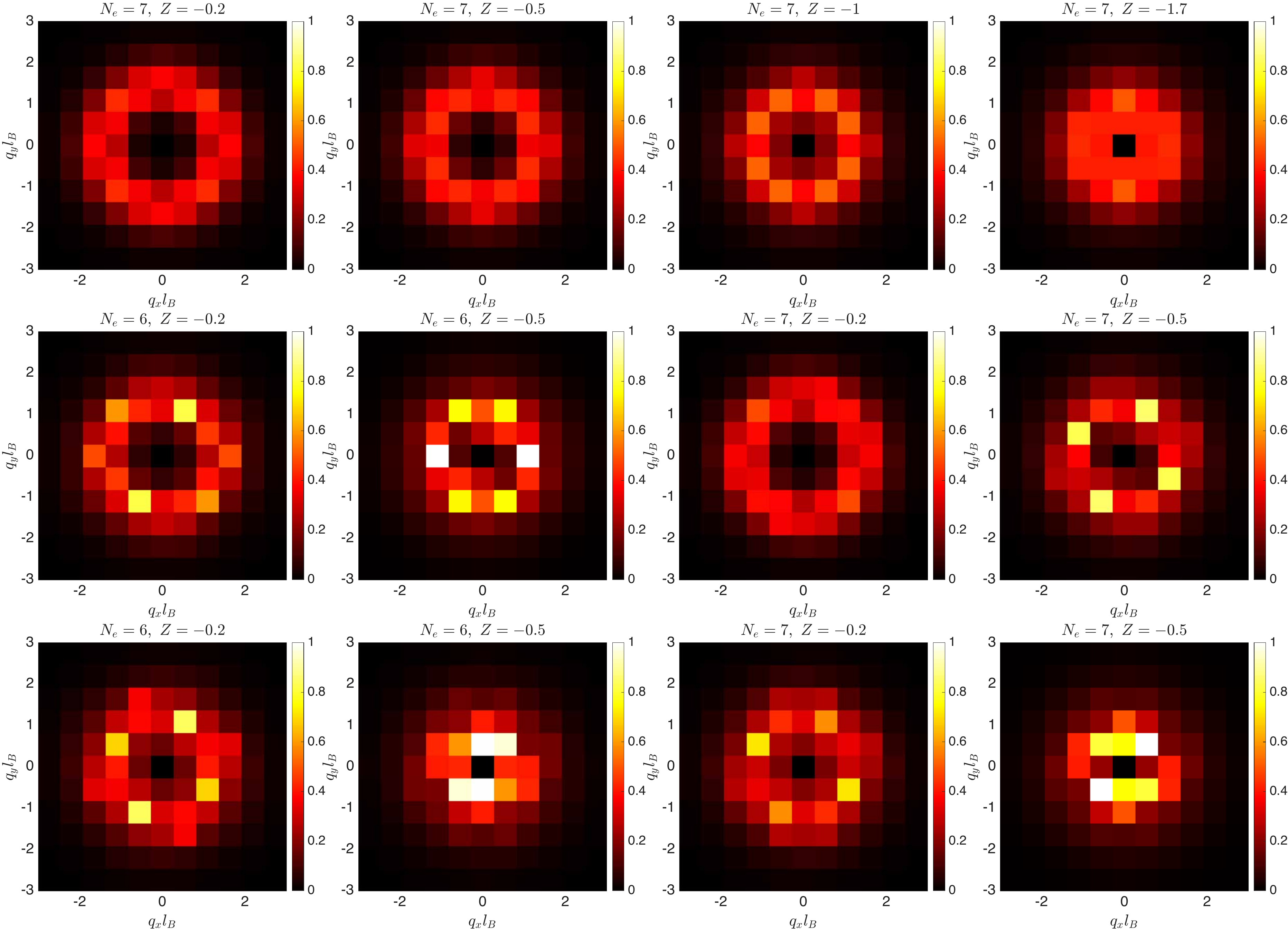}
	\caption{\label{FigSM:Coulomb} 
	Projected structure factor for the FQHL at zero temperature with charged impurities. 
	The first row is for one impurity, the second row for two impurities, the third row for six impurities. 
	Here, the parameters and the positions of the impurities are the same as those in Fig.~\ref{Fig:Coulomb} in the main text. 
	}
\end{figure*}

\begin{figure*}[!tbp]
	\center
	\includegraphics[width=0.85\textwidth]{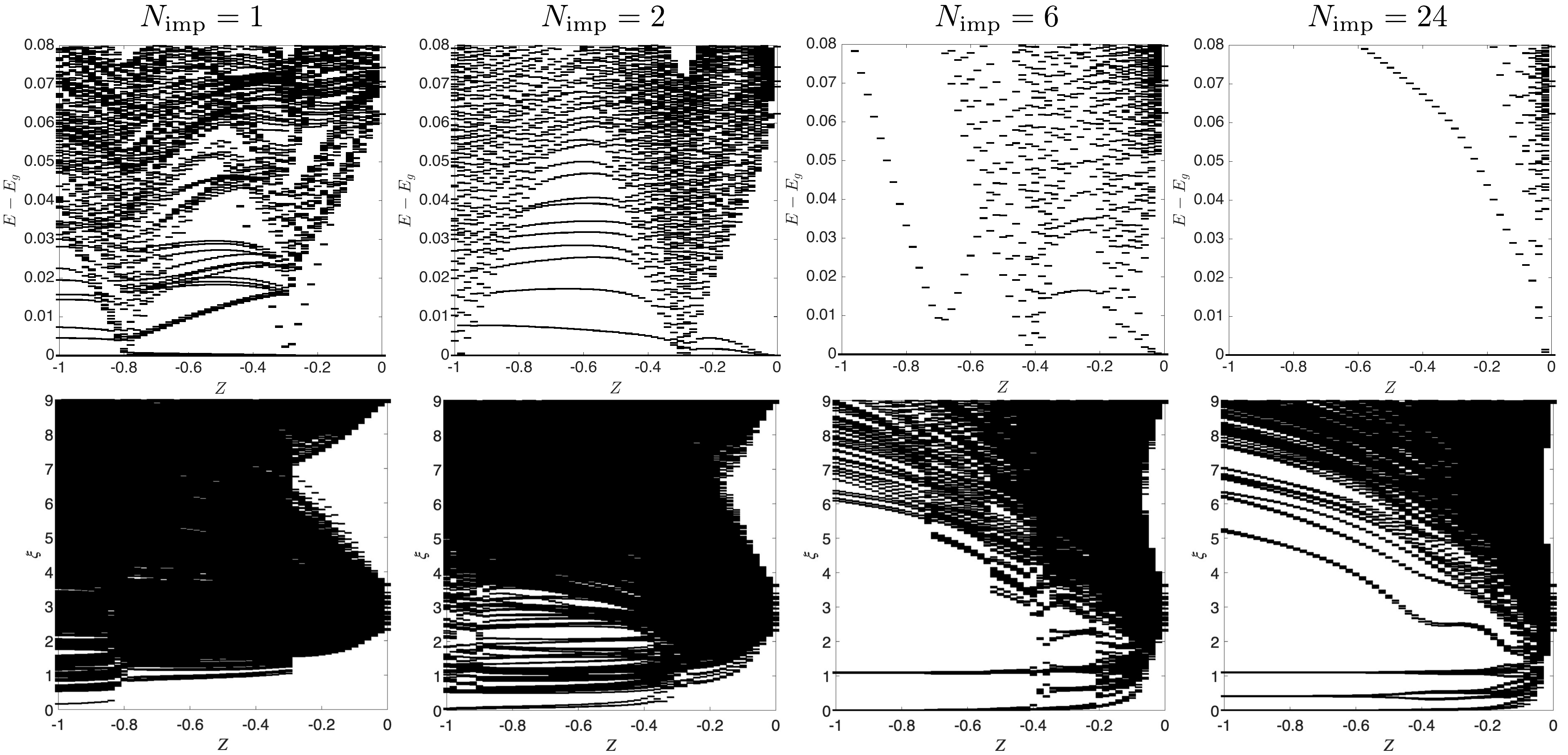}
	\caption{\label{FigSM:Coulomb_spectrum} 
	Energy and particle entanglement spectrum at the exact 1/3 filling with $N_{\text{imp}}=1,2,6,24$. Here, the parameters and the positions of the impurities are the same as those in the main text.
	}
\end{figure*}

In this section, we present the projected structure factor $\bar S(\vb* q)$ in Fig.~\ref{FigSM:Coulomb} for the FQHLs with charged impurities shown in Fig.~\ref{Fig:Coulomb} of the main text. In contrast to the random-disorder case, $\bar S(\vb* q)$ develops pronounced peaks at moderate $Z$, consistent with the crystalline order evident in the density profile. The corresponding energy spectrum and PES are shown in Fig.~\ref{FigSM:Coulomb_spectrum}. Unlike the random-disorder case, the FQH gap now closes at a finite $Z$ for every impurity density considered, and this critical $Z$ decreases as the impurity density increases.

\end{document}